\documentclass[english,twocolumn]{IEEEtran}
\usepackage{graphicx}
\usepackage{amssymb,amsbsy}
\usepackage{epstopdf}
\usepackage{amsmath,amsthm}
\usepackage{enumerate}
\usepackage{eufrak}
\usepackage{cite}
\usepackage{dsfont}
\usepackage{mathcomp}
\usepackage{supertabular}
\usepackage{makecell}
\usepackage{longtable}
\usepackage{stmaryrd}
\usepackage{url}
\usepackage{color}
\usepackage{rotating}
\usepackage{float}

\usepackage{array}
\usepackage{mathtools}
\usepackage{multicol}
\usepackage{amsmath}
\usepackage{algorithm}
\usepackage{algorithmic}
\usepackage{amssymb}
\usepackage{xcolor}
\usepackage{soul}

\usepackage[all]{xy}
\entrymodifiers={++[o][F-]}

\newcommand{\I}{\mathcal I}

\theoremstyle{definition}

\newtheorem{thm}{Theorem}
\newtheorem{cor}{Corollary}
\newtheorem{lem}{Lemma}

\newtheorem{defn}{Definition}
\newtheorem{cons}{Construction}
\newtheorem{exa}{Example}

\usepackage[T1]{fontenc}
\usepackage[latin9]{inputenc}
\usepackage{amsmath}
\usepackage{graphicx}
\usepackage{color}
\usepackage{multicol}



\usepackage{babel}



\usepackage{babel}

\begin{document}

\newcommand{\vA}{{\bf A}}

\newcommand{\vAtilde}{\widetilde{\bf A}}
\newcommand{\vB}{{\bf B}}
\newcommand{\vBtilde}{\widetilde{\bf B}}

\newcommand{\vC}{{\bf C}}
\newcommand{\vD}{{\bf D}}
\newcommand{\vH}{{\bf H}}
\newcommand{\vI}{{\bf I}}

\newcommand{\vY}{{\bf Y}}
\newcommand{\vZ}{{\bf Z}}

\newcommand{\vJ}{{\bf J}}

\newcommand{\vM}{{\bf M}}
\newcommand{\vN}{{\bf N}}
\newcommand{\vU}{{\bf U}}
\newcommand{\vV}{{\bf V}}
\newcommand{\vT}{{\bf T}}
\newcommand{\vR}{{\bf R}}
\newcommand{\vS}{{\bf S}}

\newcommand{\va}{{\bf a}}
\newcommand{\vb}{{\bf b}}
\newcommand{\vc}{{\bf c}}
\newcommand{\vd}{{\bf d}}
\newcommand{\vf}{{\bf f}}
\newcommand{\vg}{{\bf g}}

\newcommand{\ve}{{\bf e}}
\newcommand{\vh}{{\bf h}}
\newcommand{\vp}{{\bf p}}
\newcommand{\vs}{{\bf s}}

\newcommand{\vu}{{\bf u}}
\newcommand{\vv}{{\bf v}}
\newcommand{\vw}{{\bf w}}
\newcommand{\vx}{{\bf x}}
\newcommand{\vhx}{{\widehat{\bf x}}}
\newcommand{\vtx}{{\widetilde{\bf x}}}
\newcommand{\vy}{{\bf y}}
\newcommand{\vz}{{\bf z}}

\newcommand{\vj}{{\bf j}}
\newcommand{\vzero}{{\bf 0}}
\newcommand{\vone}{{\bf 1}}
\newcommand{\vbeta}{{\boldsymbol \beta}}
\newcommand{\vchi}{{\boldsymbol \chi}}

\newcommand{\dA}{\mathtt A}
\newcommand{\dT}{\mathtt T}
\newcommand{\dC}{\mathtt C}
\newcommand{\dG}{\mathtt G}

\newcommand{\tA}{\textrm A}
\newcommand{\tB}{\textrm B}
\newcommand{\A}{\mathcal A}
\newcommand{\B}{\mathcal B}
\newcommand{\C}{\mathcal C}
\newcommand{\D}{\mathcal D}
\newcommand{\E}{\mathcal E}
\newcommand{\F}{\mathcal F}
\newcommand{\G}{\mathcal G}
\newcommand{\M}{\mathcal M}
\newcommand{\HH}{\mathcal H}
\newcommand{\PP}{\mathcal P}

\newcommand{\Q}{\mathcal Q}
\newcommand{\Qb}{\bar{\mathcal Q}}
\newcommand{\Db}{{\bar{\Delta}}}

\newcommand{\pQ}{{\bf p}\mathcal Q}
\newcommand{\pQb}{{\bf p}\bar{\mathcal Q}}

\newcommand{\R}{\mathcal R}
\newcommand{\SSS}{\mathcal S}
\newcommand{\U}{\mathcal U}
\newcommand{\V}{\mathcal V}
\newcommand{\Y}{\mathcal Y}
\newcommand{\Z}{\mathcal Z}

\newcommand{\Pg}{{{\mathcal P}_{\rm gram}}}
\newcommand{\Pgint}{{{\mathcal P}^\circ_{\rm gram}}}
\newcommand{\Pgrc}{{{\mathcal P}_{\rm GRC}}}
\newcommand{\Pgrcint}{{{\mathcal P}^\circ_{\rm GRC}}}
\newcommand{\Pint}{{{\mathcal P}^\circ}}
\newcommand{\Ag}{{\bf A}_{\rm gram}}

\newcommand{\CC}{\mathbb C} 
\newcommand{\RR}{\mathbb R}
\newcommand{\ZZ}{\mathbb Z}
\newcommand{\FF}{\mathbb F}
\newcommand{\KK}{\mathbb K}

\newcommand{\Fnd}{\FF_q^{n^{\otimes d}}}
\newcommand{\Knd}{\KK^{n^{\otimes d}}}

\newcommand{\ceiling}[1]{\left\lceil{#1}\right\rceil}
\newcommand{\floor}[1]{\left\lfloor{#1}\right\rfloor}
\newcommand{\bbracket}[1]{\left\llbracket{#1}\right\rrbracket}

\newcommand{\inprod}[1]{\left\langle{#1}\right \rangle}


\newcommand{\beas}{\begin{eqnarray*}} 
\newcommand{\eeas}{\end{eqnarray*}} 

\newcommand{\bm}[1]{{\mbox{\boldmath $#1$}}} 

\newcommand{\sizeof}[1]{\left\lvert{#1}\right\rvert}
\newcommand{\wt}{{\rm wt}} 
\newcommand{\supp}{{\rm supp}} 
\newcommand{\dg}{d_{\rm gram}} 
\newcommand{\da}{d_{\rm asym}} 
\newcommand{\dist}{{\rm dist}} 
\newcommand{\ssyn}{s_{\rm syn}}
\newcommand{\sseq}{s_{\rm seq}}
\newcommand{\nullplus}{{\rm Null}_{>\vzero}}

\newcommand{\tworow}[2]{\genfrac{}{}{0pt}{}{#1}{#2}}
\newcommand{\qbinom}[2]{\left[ {#1}\atop{#2}\right]_q}

\newcommand{\Lovasz}{Lov\'{a}sz }
\newcommand{\etal}{\emph{et al.}}

\newcommand{\todo}{{\color{red} (TODO) }}


\title{Mutually Uncorrelated Primers for DNA-Based Data Storage}

\author{
  \IEEEauthorblockN{
    S. M. Hossein~Tabatabaei Yazdi\IEEEauthorrefmark{1}, Han Mao Kiah\IEEEauthorrefmark{2}, Ryan Gabrys\IEEEauthorrefmark{1}~and~
    Olgica~Milenkovic\IEEEauthorrefmark{1}}
  {\normalsize
    \begin{tabular}{ccc}
      \IEEEauthorrefmark{1}ECE Department, University of Illinois, Urbana-Champaign~~ &
      \IEEEauthorrefmark{2}SPMS, Nanyang Technological University, Singapore \\
    \end{tabular}}\vspace{-3ex}
    \thanks{Parts of the work were presented at ISIT 2016, Barcelona, Spain. The research is supported in part by the NSF CCF 16-18366 grant.}
}
\maketitle
\begin{abstract}
We introduce the notion of weakly mutually uncorrelated (WMU) sequences, 
motivated by applications in DNA-based data storage systems and for synchronization of communication devices. 
WMU sequences are characterized by the property that no sufficiently long suffix of one sequence is 
the prefix of the same or another sequence. WMU sequences used for primer design in DNA-based data storage systems are also required to be at large mutual Hamming distance from each other, have balanced compositions of symbols, and avoid primer-dimer byproducts. We derive bounds on the size of WMU and various constrained WMU codes and present a number of constructions for balanced, error-correcting, primer-dimer free WMU codes using Dyck paths, prefix-synchronized and cyclic codes. 
\end{abstract}
\IEEEpeerreviewmaketitle
\section{Introduction}
Mutually uncorrelated (MU) codes are a class of fixed length block codes in which no proper prefix of one codesequence is a suffix of the same or another codesequence.
MU codes have been extensively studied in the coding theory and combinatorics literature under a variety of names. Levenshtein introduced the codes in 1964 under the name `strongly regular codes'~\cite{levenshtein1964decoding}, 
and suggested that the codes be used for synchronization. For frame synchronization applications described by van Wijngaarden and Willink in~\cite{de2000frame}, 
Baji\'c and Stojanovi\'c~\cite{bajic2004distributed} rediscovered MU codes, and studied them under the name of `cross-bifix-free' codes. Constructions and bounds on the size of MU codes were also reported in a number of recent contributions~\cite{bilotta2012new, blackburn2013non}. In particular, Blackburn~\cite{blackburn2013non} analyzed these sequences under the name of `non-overlapping codes', and  provided a simple construction for a class of codes with optimal cardinality. 

MU codes have recently found new applications in DNA-based data storage~\cite{church2012next,goldman2013towards}: In this setting, Yazdi~\etal~\cite{yazdi2015rewritable,yazdi2017portable} developed a new, random-access and rewritable DNA-based data storage architecture that uses MU address sequences that allow selective access to encoded DNA blocks via Polymerase Chain Reaction (PCR) amplification with primers complementary to the address sequences. In a nutshell, DNA information-bearing sequences are prepended with address sequences used to access strings of interest via PCR amplification. To jump start the amplification process, one needs to `inject' complements of the sequences into the storage system, and those complementary sequences are referred to as DNA primers. Primers attach themselves to the user-selected address strings and initiate the amplification reaction. In order to ensure accurate selection and avoid expensive postprocessing, the information sequences following the address are required to avoid sequences that resemble the addresses, thereby imposing a special coding constraint that may be met through the use of MU addresses. In addition, the addressing scheme based on MU codes may be used in conjunction with other specialized DNA-based data storage codes like the ones outlined in~\cite{kiah2015codes,gabrys2016codes,gabrys2017asymmetric}. Detailed descriptions of implementations of DNA-based data storage systems and their underlying synthetic biology principles are beyond the scope of this paper; the interested reader is referred to~\cite{yazdi2015dna} for a discussion of system components and constraints. 

The goal of this work is to generalize the family of MU codes by introducing weakly mutually uncorrelated (WMU) codes. WMU codes are block codes in which no \emph{sufficiently long  
prefix} of one codesequence is a suffix of the same or another codesequence. In contrast, MU codes 
prohibit suffix-prefix matches of any length. This relaxation of prefix-suffix constraints was motivated in~\cite{yazdi2015rewritable}, with the purpose of improving code rates and allowing for increased precision DNA fragment assembly and selective addressing. A discussion of the utility of WMU codes in DNA-based data storage may be found in the overview paper~\cite{yazdi2015dna, yazdi2017portable} and the paper describing recent practical implementations of portable DNA-based data storage systems which make use of WMU codes~\cite{yazdi2016portable}.

Here, we are concerned with determining bounds on the size of specialized WMU codes and efficient WMU code constructions. Of interest are both binary and quaternary WMU codes, as the former may be used to construct the latter, while the latter class may be adapted for encoding over the four letter DNA alphabet $\{{\tt A,T,C,G}\}$. Our contributions include bounds on the largest size of unconstrained and constrained WMU codes, constructions of WMU codes that meet the derived upper bounds as well as results on several important constrained versions of WMU codes: Error-correcting WMU codes, balanced WMU codes, balanced error-correcting WMU codes, and WMU codes that avoid primer-dimer byproducts. The aforementioned constraints arise due to the following practical considerations.

A binary sequence is called balanced if half of its symbols are zero. On the other hand, a DNA sequence is termed balanced if it has a $50\%$ GC content (i.e., if $50\%$ of the symbols in the sequence are either $\tt G$ or $\tt C$). Balanced DNA sequences are more stable than DNA sequences with lower or higher GC content and they have lower sequencing error-rates. Balanced DNA sequences are also easier to synthesize than unbalanced sequences~\cite{yakovchuk2006base}. In addition, WMU codes at large Hamming distance limit the probability of erroneous codesequence selection dues to address errors. When referring to primer dimer (PD) issues~\cite{vallone2004autodimer}, we consider potential problems that may arise during random access when two primers used for selection bond to each other, thereby prohibiting amplification of either of the two corresponding information-bearing sequences. PD byproducts can be eliminated by restricting the WMU codes to avoid simultaneous presence of long substrings and their complements in the codesequences.

The paper is organized as follows. Section~\ref{sec:roadmap} contains an overview of the topics and results discussed in the paper and some formal definitions needed to follow the material in subsequent sections. In Section~\ref{sec:basics} we review MU and introduce WMU codes, and derive bounds on the maximum size of the latter family of combinatorial objects. In addition, we outline a construction of WMU codes that meets the derived upper bound. We also describe a construction that uses binary MU component codes and other constrained codes in order to obtain families of WMU codes that obey different combinations of primer constraints. In Section~\ref{sec:wmuerror} we describe constructions for error-correcting WMU codes, while in Section~\ref{sec:wmubalanced} we discuss balanced WMU codes. Primer-dimer constraints are discussed in Section~\ref{sec:pd}. Our main results are presented in Section~\ref{sec:all}, where we first propose to use cyclic codes to devise WMU codes that are both balanced and have error correcting capabilities. We then proceed to improve the cyclic code construction in terms of coding rate through decoupled constrained and error-correcting coding for binary strings. In this setting, we use DC-balanced codes~\cite{immink2004codes}. Encoding of information with WMU address codes is described in Section~\ref{sec:dna_code}.

\section{Roadmap of Approaches and Results} \label{sec:roadmap}

Throughout the paper we use the following notation: $\FF_q$ stands for a finite field of order $q \geq 2$. 
Two cases of special interest are $q=2$ and $q=4$. In the latter case,
we tacitly identify the elements of $\FF_4$ with the four letters of the DNA code alphabet, $\{{\mathtt{A,T,C,G}\}}$.
We let $\va=\left(a_{1},\ldots,a_{n}\right) \in\FF_q^n $ stand for a sequence of length $n$ over $\FF_q$, and let $\mathbf{a}_{i}^{j}$,
$1\leq i,j\leq n$, stand for a substring of $\mathbf{a}$ starting
at position $i$ and ending at position $j$, i.e., 
\[
\mathbf{a}_{i}^{j}=\begin{cases}
\left(a_{i},\ldots,a_{j}\right) & i\leq j\\
\left(a_{i},a_{i-1},\ldots,a_{j}\right) & i>j.
\end{cases}
\]
Moreover, for two arbitrary sequences $\va \in\FF_q^n,\vb \in\FF_q^m,$ we use $\va \vb = (a_{1},\ldots,a_{n}, b_{1},\ldots,b_{m})$ to denote a sequence of length $n+m$ generated by appending $\vb$ to the right of $\va$. Thus, $\va^{l}$ stands for a sequence of length $ln$ comprising $l$ consecutive copies of the sequence $\va$. 

We say that a sequence $\bar{\mathbf{a}}=\left(\bar{a_{1}},\ldots,\bar{a_{n}}\right)\in\mathbb{F}_{q}^{n}$
represents the complement of sequence $\mathbf{a}\in\mathbb{F}_{q}^{n}$ if:
\begin{itemize}
\item For $q=2$, and $1\leq i\leq n,$
\end{itemize}
\begin{equation}
\bar{a}_{i}=\begin{cases}
1 & \textrm{if }a_{i}=0,\\
0 & \textrm{if }a_{i}=1;
\end{cases}\label{eq:comp_2}
\end{equation}
\begin{itemize}
\item For $q=4$, and $1\leq i\leq n,$
\end{itemize}
\begin{equation}
\bar{a}_{i}=\begin{cases}
\mathtt{T} & \textrm{if }a_{i}=\mathtt{A},\\
\mathtt{A} & \textrm{if }a_{i}=\mathtt{T},\\
\mathtt{G} & \textrm{if }a_{i}=\mathtt{C},\\
\mathtt{C} & \textrm{if }a_{i}=\mathtt{G}.
\end{cases}\label{eq:comp_4}
\end{equation}
The notion of complement used for $\mathbb{F}_{4}$ is often referred to as the Watson-Crick (W-C) complement.

In this work, we define an (address) code $\C$ of length $n$ as a collection of sequences from $\FF_q^n$, for $q \in \{{2,4\}}$, satisfying a set of specific combinatorial constraints described below.

The goal is to describe new constructions for address sequences used for DNA-based data storage.
Address sequences should enable reliable access to desired information content. This is accomplished by making the addresses as distinguishable from each other as possible via a simple minimum Hamming distance constraint; recall that the Hamming distance $d_H$ between any two sequences of length $n$, $\va=(a_1,\ldots,a_n)$ and $\vb=(b_1,\ldots,b_n)$, over some finite alphabet $\mathcal{A}$ equals 
$$ d_H(\va,\vb)=\sum_{i=1}^n\, \mathds{1}(a_i \neq b_i),$$
where $\mathds{1}(\cdot)$ stands for the indicator function. 
One may also use the Levenshtein distance instead, as discussed in the context of MU codes in~\cite{levy2017mutually}.  

Access to desired sequences is accomplished by exponentially amplifying them within the pool of all sequences via addition 
of primer sequences corresponding to the W-C complement of their addresses. As primers have to be synthesized, they need to satisfy constraints that enable simplified synthesis, such as having a balanced GC-content, formally defined for a sequence $\va$ over $\FF_4^n$ as $\sum_{i=1}^n \mathds{1}(a_i \in \{ \mathtt{G,C}\}) = \frac{n}{2}$. This constraint directly translates to a balancing property for the address sequences. Furthermore, as one may require simultaneous amplification of multiple sequences, multiple primers need to be added in which case it is undesirable for different pairs of primers to bond to each other via W-C complementarity. The PD byproducts of this binding may be significantly reduced if one imposes an additional PD constraint on the primers, and hence on the address sequences, as defined below.    

\begin{defn} A set of sequences $\C \subseteq\mathbb{F}_{q}^{n}$, for $q\in\left\{ 2,4\right\} $, is said to avoid primer dimer (APD) byproducts of effective length $f$ if substrings of sequences in $\C$ with length $\geq f$ cannot hybridize with each other in the forward or the reverse direction. More precisely,
we say that $\C$ is an $f$-APD code if for any two sequences $\mathbf{a},\mathbf{b}\in\C$, not necessarily distinct, and $1\leq i,j\leq n+1-f,$ we have $\mathbf{\bar{a}}_{i}^{f+i-1}\neq \mathbf{b}_{j}^{f+j-1},\mathbf{b}_{f+j-1}^{j}$. We refer to the sequence $\mathbf{b}_{f+j-1}^{j}$ as the \emph{reverse} of the sequence $\mathbf{b}_{j}^{f+j-1}$.
\end{defn}
For practical reasons, we only focus on the parameter regime $f=\Theta\left(n\right)$, as only sufficiently long complementary sequences may bond with each other. Furthermore, we defer the study of the related problem of secondary structure formation~\cite{milenkovic2006design,milenkovic2005dna} to future work.

In certain DNA-based data storage systems, one may be interested in restricting the address sequences by imposing only one or two of the above constraints. For example, if the addresses are relatively short ($\leq 10$), one may dispose of the requirement to make the sequences balanced, as short sequences are significantly easier to synthesize than longer ones. If one allows for postprocessing of the readouts, then the Hamming distance constraint may be relaxed or completely removed. It is for this reason that we also consider a more general class of code constructions that accommodate only a subset of the three previously described constraints. 

By far the most important constraint imposed on the address sequences is that they enable a simple construction of information-bearing sequences (assumed to be of length $N>>n$) that do not contain any of the address sequences of length $n$ as substrings. It is in this context of forbidden substring coding that MU codes were introduced in~\cite{gilbert1960synchronization,guibas1978maximal}. WMU codes may be used in the same setting, but they are less restrictive than MU codes, and therefore allow for larger codebooks. This is why our main results pertain to constructions of WMU codes with various subsets of primer constraints, and we formally define and discuss these codes in the next section. For some related questions pertaining to MU codes, the interested reader is referred to~\cite{levy2017mutually}.
 
\section{MU and WMU Codes: Definitions, Bounds and Constructions} \label{sec:basics}
For simplicity of notation, we adopt the following naming convention for codes:
If a code $\C\subseteq\FF_q^n$ has properties $\textrm{Property}_1, \textrm{Property}_2,\ldots, \textrm{Property}_s$, then we say that $\C$  is a $\mathtt{Property_1\_Property_2\_\ldots,Property_s\_q\_n}$ code, and use the previous designation in the subscript. 

\subsection{Mutually Uncorrelated Codes}
We say that a sequence $\va=\left(a_{1},\ldots,a_{n}\right) \in\FF_q^n $ is self-uncorrelated if no proper prefix of $\va$ matches its suffix, i.e., if $\left(a_{1},\ldots,a_{i}\right) \neq \left(a_{n-i+1},\ldots,a_{n}\right)$, for all $ 1\leq i < n$. This definition may be extended to a set of sequences as follows: Two not necessarily distinct sequences $\va , \vb \in \FF_q^n$ are said to be mutually uncorrelated if no proper prefix of $\va$ appears as a suffix of $\vb$ and vice versa. We say that $\C\subseteq\FF_q^n$ is a mutually uncorrelated (MU) code if any two not necessarily distinct codesequences in $\C$ are mutually uncorrelated.

The maximum cardinality of MU codes was determined up to a constant factor by Blackburn~\cite[Theorem 8]{blackburn2013non}. For completeness, 
we state the modified version of this result for alphabet size $q \in \{2,4\}$ below

\begin{thm}
Let $A_{\mathtt{MU\_q\_n}}$ denote the maximum size of a $\mathtt{MU\_q\_n}$ code, with $n\geq 2$ and $q \in \{2,4\}$. Then
\label{thm:M1}
\[
c_q \frac{ q^n}{n} \leq A_{\mathtt{MU\_q\_n}} \le \frac{ q^n}{2n}
\]
where $c_q = \frac{(q-1)^2 (2q-1)}{4q^4},$ which for $q=2$ and $q=4$ equal $c_2 = 0.04688$ and $c_4 =0.06152$, respectively. 
\end{thm}
We also briefly outline two known constructions of MU codes, along with a new and simple construction for error-correcting MU codes that will be used in our subsequent derivations.

Bilotta \etal{}~\cite{bilotta2012new} described an elegant construction for MU codes based on well-known combinatorial objects termed Dyck sequences. A Dyck sequence of length $n$ is a binary sequence composed of $\frac{n}{2}$ zeros and $\frac{n}{2}$ ones such that no prefix of the sequence has more zeros 
than ones. By definition, a Dyck sequence is balanced and it necessarily starts with a one and ends with a zero. The number of Dyck word of length $n$ is the $\frac{n}{2}$-th Catalan number, equal to $\frac{2}{n+2}\binom{n}{\frac{n}{2}}$.
\begin{cons}\label{cons:BAL_MU_2_n} ($\mathtt{BAL\_MU\_2\_n}$ Codes)
Consider a set $\D$ of Dyck sequences of length $n-2$ and define the following set of sequences of length $n$,
\[\C = \{1\va 0: \va\in\D\}.\]
\end{cons}
It is straightforward to show that $\C$ is balanced and MU code. 
Size of $\C$ is also equal to  $\frac{n-2}{2}$-th Catalan number, or $|\C| = \frac{1}{2(n-1)}\binom{n}{\frac{n}{2}}$.

An important observation is that MU codes constructed using Dyck sequences are inherently balanced, as they contain $\frac{n}{2}$ ones and $\frac{n}{2}$ zeros. The balancing property also carries over to all prefixes of certain subsets of Dyke sequences. To see this, recall that a Dyck sequence has \emph{height} at most $D$ if for any prefix of the sequence, the difference between the number of ones and the number of zeros 
is at most $D$. Hence, the disbalance of any prefix of a Dyck sequence of height 
$D$ is at most $D$. Let Dyck$(n,D)$ denote the number of Dyck sequences of length $2n$ and height at most $D$. For fixed values of $D$, de Bruijn \etal{} \cite{bruijn1972average} proved that 
\begin{equation} \label{eq:d-D}
{\rm Dyck}(n,D)\sim\frac{4^n}{D+1}\tan^2 \left(\frac{\pi}{D+1}\right) \cos^{2n} \left(\frac{\pi}{D+1} \right).
\end{equation}
Here, $f(n)\sim g(n)$ is used to denote the following asymptotic relation $\lim_{m\to\infty} f(n)/g(n)=1$. 

Bilotta's construction also produces nearly prefix-balanced MU codes, provided that one restricts his/her attention to subsets of sequences with small disbalance $D$; equation~\ref{eq:d-D} establishes the existence of large subsets of Dyck sequences with small disbalance. By mapping $0$ and $1$ to $\{\mathtt{A},\mathtt{T}\}$ and $\{\mathtt{C},\mathtt{G}\}$, respectively, one may enforce a similar GC balancing constraint on DNA MU codes.

The next construction of MU codes was proposed by Levenshtein~\cite{levenshtein1964decoding} and Gilbert~\cite{gilbert1960synchronization}. 
\begin{cons}\label{cons:MU_q_n} ($\mathtt{MU\_q\_n}$ Codes)
Let $n \geq 2$ and $1\leq \ell \leq n-1$, be two integers and let $\C \subseteq \FF_q^n$ be the set of all sequences $\va=\left(a_{1},\ldots,a_{n}\right)$ such that
\begin{itemize}
\item The sequence $\va$ starts with $\ell$ consecutive zeros, i.e., $\va_{1}^\ell = 0^\ell$.
\item It holds that $a_{\ell +1},a_{n} \neq 0$.
\item The subsequence $\va_{\ell +2}^{n-1}$ does not contain $\ell$ consecutive zeros as a subsequence.
\end{itemize}
\end{cons}
Then, $\C$ is an MU code.
Blackburn~\cite[Lemma 3]{blackburn2013non} showed that when $\ell = \left\lceil \log_q 2n \right\rceil$ and $n \geq 2\ell + 2$ the above construction is optimal. His proof relies on
the observation that the number of strings $\va_{\ell +2}^{n-1}$ that do not contain $\ell$ consecutive zeros as a subsequence exceeds $\frac{ \left(q-1\right)^{2}\left(2q-1\right)}{4nq^4}q^n$, thereby establishing the lower bound of Theorem \ref{thm:M1}. The aforementioned result is a simple consequence of the following lemma.
\begin{lem}
\label{lem:avoid_string} 
The number of $q$-ary sequences of length $n$ that avoid $t$ specified sequences in $\FF_q^{n_s}$ as substrings is greater than $q^{n} (1-\frac{n  t}{q^{n_s}})$.
\end{lem}
\begin{IEEEproof}
The result obviously holds for $n \leq n_s$. If $n \ge n_s$, then the number of bad strings, i.e., $q$-ary strings of length $n$ that contain at least one of the specified $t$ strings as a substring, is bounded from above by:
\begin{align*}
\# \textrm{bad strings} & \le (n - n_s  + 1) t q^{n - n_s}\\
& \le   n t  q^{n - n_s}.
\end{align*}
Hence, the number of good sequences, i.e., the number of $q$-ary sequences of length $n$ that avoid $t$ specified strings in $\FF_q^{n_s}$ as substrings, is bounded from below by
\begin{align*}
\# \textrm{good strings} & \ge q^{n} - \# \textrm{bad strings}\\
& \ge q^{n} (1-\frac{n t}{q^{n_s}}).
\end{align*}
\end{IEEEproof}


It is straightforward to modify Construction \ref{cons:MU_q_n} so as to incorporate error-correcting redundancy. Our constructive approach to this problem is outlined in what follows.
\begin{cons}\label{cons:Error-Correcting_MU_2_n} ($\mathtt{Error-Correcting\_MU\_2\_n}$ Codes)
Fix two positive integers $t$ and $\ell$ and consider a binary $(n_H,s,d)$ code $\C_H$
of length $n_H=t(\ell-1)$, dimension $s$, and Hamming distance $d$.
For each codesequence $\vb \in \C_H$,
we map $\vb$ to a sequence of length $n=(t+1)\ell+1$ given by
\[\va(\vb)=0^\ell 1\vb^{\ell -1}_{1}1\vb^{2(\ell -1)}_{\ell}1\cdots\vb^{t(\ell -1)}_{(t-1)(\ell -1)+1}1.\]
Let $\C_{\rm parse}\triangleq \{\va(\vb): \vb\in \C_H\}$.
\end{cons}
It is easy to verify that $|\C_{parse}| = |\C_H|$, and that the code $\C_{parse}$ has the same minimum Hamming distance as $\C_H$, i.e., $d(\C_{parse})=d(\C_H)$. As $n_H=t(\ell-1)$, we also have $\C_{parse} \subseteq \left\{ 0,1 \right\}^n$, where $n=(t+1)\ell+1$. 
In addition, the parsing code $\C_{parse}$ is an MU code, since it satisfies all the constraints required by Construction~\ref{cons:MU_q_n}. 
To determine the largest asymptotic size of a parsing code, we recall the Gilbert-Varshamov bound.
\begin{thm}\label{thm:GV}
(Asymptotic Gilbert-Varshamov bound \cite{gilbert1952comparison,varshamov1957estimate})
For any two positive integers $n$ and $d \leq \frac{n}{2},$ there exists a block code $\C \subseteq \left \{ 0,1 \right \}^n$ of minimum Hamming distance $d$ with normalized rate  
\begin{equation*}
R(\C) \geq 1 - h\left(\frac{d}{n}\right) -o(1),
\end{equation*}
where $h(\cdot)$ is the binary entropy function, i.e., $h(x)=x \log_2 \frac{1}{x} + (1-x) \log_2 \frac{1}{1-x}$, for $0 \leq x \leq 1$.
\end{thm}
Recall that the parameters $s$ (dimension) and $d$ (minimum Hamming distance) of the codes $\mathcal{C}_{H}$ and $\mathcal{C}_{parse}$ are identical. Their lengths, $n_H$ and $n$, respectively, equal $n_H=t\left(\ell-1\right)$ and $n=\left(t+1\right)\ell+1$, where $t,\ell$ are positive integers.
We next aim to optimize the parameters of the \emph{parsing code} for fixed $s$ and fixed $n$, which amounts to maximizing $d$. Since $d$ is equal to the corresponding minimum distance of the code $\mathcal{C}_H$, and both codes have the same dimension $s$, in order to maximize $d$ we maximize $n_H$ under the constraint that $n$ is fixed. More precisely, we optimize the choice of $\ell,t$ and then use the resulting parameters in the Gilbert-Varshamov lower bound.

To maximize $n_{H}=t\left(\ell-1\right)$ given $n=\left(t+1\right)\ell+1$
and $t,\ell\geq1$, we write 
$$n_{H}= n-(\ell+t+1) \leq n-2\sqrt{\ell(t+1)}= n-2\sqrt{n-1}.$$
Here, the inequality follows from the arithmetic and geometric mean inequality, i.e., $\frac{\ell+t+1}{2}\geq\sqrt{\ell \left(t+1\right)}$. On the other hand, it is easy to verify that this upper bound is achieved by setting $\ell=\sqrt{n-1}$ and $t=\sqrt{n-1}-1$. Hence, the maximum value of $n_{H}$ is $n_{H}^{*}=n-2\sqrt{n-1}$.

By using a code $\C_H$ with parameters $[n^{\ast}_{H}, s, d]$ as specified by the GV bound, where $d \leq \frac{n^{\ast}_{H}}{2}$ and $s = n^{\ast}_{H} \, (1-h(\frac{d}{n^{\ast}_{H}})-o(1))$, we obtain an error-correcting MU code $\C_{parse}$ with parameters $[n^{\ast}_{H}+2\sqrt{n^{\ast}_{H}+2\sqrt{n^{\ast}_{H}-1}-1},n^{\ast}_{H} \, (1-h(\frac{d}{n^{\ast}_{H}})-o(1)),d]$.

\subsection{Weakly Mutually Uncorrelated Codes: Definitions, Bounds and Constructions}
The notion of mutual uncorrelatedness may be relaxed by requiring that 
only sufficiently long prefixes of one sequence do not match sufficiently long suffixes of the same or another sequence. A formal definition of this property is given next. 
\begin{defn}
Let $\C\subseteq\FF_q^n$ and $ 1\leq \kappa < n$.
We say that $\C$ is a $\kappa$-weakly mutually uncorrelated ($\kappa$-WMU) code if 
no proper prefix of length $l$, for all $l \geq \kappa$, of a codesequence in $\C$ appears as a suffix of another codesequence, including itself.
\end{defn}
Our first result pertains to the size of the largest WMU code.

\begin{thm}
\label{thm:M_2}Let $A_{\mathtt{\kappa-WMU\_q\_n}}$ denote the maximum size of a $\kappa$-WMU code over $\mathbb{F}_{q}^{n}$, for $1 \leq \kappa < n$ and $q \in \{2,4\}$. 
Then, 
\[
c_q \, \frac{q^{n}}{n-\kappa+1} \leq A_{\mathtt{\kappa-WMU\_q\_n}} \leq \frac{q^{n}}{n-\kappa+1},
\]
where the constant $c_q$ is as described in Theorem \ref{thm:M1}.
\end{thm}
\begin{IEEEproof}
\label{proof_M_2_bounds}To prove the upper bound, we use an approach first suggested by Blackburn in~\cite[Theorem 1]{blackburn2013non}, for the purpose of analyzing MU codes. 
Assume
that $\C \subseteq \mathbb{F}_{q}^{n}$ is a $\kappa$-WMU code.
Let $L=\left( n+1\right)\left(n-\kappa+1\right)-1$, and consider the set $X$ of pairs $\left(\va ,i\right),$ where $i\in\left\{ 1,\ldots,L\right\} $, and where $\va \in\mathbb{F}_{q}^{L}$ is such that the (possibly cyclically wrapped) substrings of $\va$ of length $n$ starting at position $i$ belongs to $\C$. Note that our choice of the parameter $L$ is governed by the overlap length $\kappa$.

Clearly, $\left|X\right|=L\left|\C\right|q^{L-n}$, since there are
$L$ different possibilities for the index $i$, $\left|\C\right|$ possibilities for
the string starting at position $i$ of $\va$, and $q^{L-n}$ choices
for the remaining $L-n\geq 0$ symbols in $\va$. Moreover, if $\left(\va,i\right)\in X,$
then $\left(\va,j\right)\notin X$ for $j\in\left\{ i\pm1,\ldots,i\pm n-\kappa\right\} _{\textrm{mod }L}$ due to the weak mutual uncorrelatedness property.
Hence, for a fixed string $\va\in\mathbb{F}_{q}^{L}$, there
are at most $\left\lfloor \frac{L}{n-\kappa+1}\right\rfloor$ different
pairs $\left(\va,i_{1}\right),\ldots,\left(\va,i_{\left\lfloor \frac{L}{n-\kappa+1}\right\rfloor}\right)\in X$. This implies that 
$$\left|X\right|\leq \left\lfloor \frac{L}{n-\kappa+1}\right\rfloor q^{L}.$$ 
Combining the two derived constraints on the size of $X$, we obtain
$$\left|X\right|=L\left|\C\right|q^{L-n}\leq\left\lfloor \frac{L}{n-\kappa+1}\right\rfloor q^{L}.$$
Therefore, $ \left|\C\right| \leq \frac{q^{n}}{n-\kappa+1}$.

To prove the lower bound, we describe a simple WMU code construction, outlined in Construction~\ref{cons:WMU_q_n}. 
\begin{cons}\label{cons:WMU_q_n}($\mathtt{\kappa-WMU\_q\_n}$ Codes)
Let $\kappa,n$ be two integers such that $1\leq \kappa \leq n$. A $\kappa$-WMU code $\C \in \FF_q^n$ may be constructed using a simple concatenation of the form
 $\C=\left\{ \va \vb \mid \va \in \C_{1}, \vb \in \C_{2}\right\} $, where $\C_1\subseteq\FF_q^{n-\kappa +1}$ is an MU code, and $\C_2 \subseteq\FF_q^{\kappa -1}$ is unconstrained. 
\end{cons}
It is easy to verify that $\C$ is an $\kappa$-WMU code with $\left|\C_1\right| \, \left|\C_2\right|$ codesequences.
Let  $\C_2 = \mathbb{F}_{q}^{\kappa -1}$ and let $\C_1 \subseteq\FF_q^{n-\kappa +1}$ be the largest MU code of size $A_{\mathtt{MU\_q\_n-\kappa+1}}$.
Then, $\left|\C\right| = q^{\kappa -1} \, A_{\mathtt{MU\_q\_n-\kappa+1}}$. The claimed lower bound now follows from the lower bound of Theorem~\ref{thm:M1}, establishing that $\left|\C\right| \geq c_q \, \frac{q^{n}}{n-\kappa+1}.$ 
\end{IEEEproof}
As described in the Introduction, $\kappa$-WMU codes used in DNA-based storage applications are required to satisfy a number of additional combinatorial constraints in order to be used as blocks addresses. These include the error-correcting, balancing and primer dimer constraints. Balancing and error-correcting properties of codesequences have been studied in great depth, but not in conjunction with MU or WMU codes. The primer dimer constraint has not been previously considered in the literature.

In what follows, we show that all the above constraints can be imposed on $\kappa$-WMU codes via a simple \emph{decoupled binary code construction}. To this end, let us introduce a mapping $\Psi$ as follows. For any two binary
sequences $\va=\left(a_{1},\ldots,a_{n}\right), \vb=\left(b_{1},\ldots,b_{n}\right)\in\left\{ 0,1\right\} ^{n}$, $\Psi\left(\va,\vb\right):\left\{ 0,1\right\} ^{n}\times\left\{ 0,1\right\} ^{n}\rightarrow\left\{ \mathtt{A},\mathtt{T},\mathtt{C},\mathtt{G}\right\} ^{n}$
is an encoding function that maps the pair $\va, \vb$ to a DNA string $\vc=\left(c_{1},\ldots,c_{n}\right)\in\left\{ \mathtt{A},\mathtt{T},\mathtt{C},\mathtt{G}\right\} ^{n}$, according to the following rule:
\begin{equation}
\textrm{for }1\leq i\leq n,\: c_{i}=\begin{cases}
\mathtt{A} & \textrm{if }\left(a_{i}, b_{i}\right)=\left(0,0\right)\\
\mathtt{T} & \textrm{if }\left(a_{i}, b_{i}\right)=\left(0,1\right)\\
\mathtt{C} & \textrm{if }\left(a_{i}, b_{i}\right)=\left(1,0\right)\\
\mathtt{G} & \textrm{if }\left(a_{i}, b_{i}\right)=\left(1,1\right).
\end{cases}\label{eq:mapping}
\end{equation}
Clearly, $\Psi$ is a bijection and $\Psi ( \va, \vb ) \Psi ( \vc, \vd ) =\Psi ( \va \vc, \vb \vd)$.
The next lemma lists a number of useful properties of $\Psi$.
\begin{lem}
\label{lem:DNA_mapping_properties} Suppose that $\C_{1},\C_{2}\subseteq\left\{ 0,1\right\} ^{n}$
are two binary block codes of length $n$. Encode pairs of codesequences $\left( \va, \vb \right)\in \C_{1}\times \C_{2}$ into a code $\C=\left\{ \Psi\left( \va , \vb \right)\mid \va \in \C_{1}, \vb\in \C_{2}\right\} $. Then: 
\begin{enumerate}
\item If $\C_{1}$ is balanced, then $\C$ is balanced.
\item If either $\C_{1}$ or $\C_{2}$ are $\kappa$-WMU codes, then $\C$ is also an $\kappa$-WMU code.
\item If $d_{1}$ and $d_{2}$ are the minimum Hamming distances of $\C_{1}$
and $\C_{2}$, respectively, then the minimum Hamming distance of $\C$
is at least $\min\left(d_{1},d_{2}\right)$.
\item If $\C_{2}$ is an $f$-APD code, then $\C$ is also an $f$-APD code.
\end{enumerate}
\end{lem}
\begin{IEEEproof}
\label{proof_DNA_mapping_properties}
\begin{enumerate}

\item Any $\vc\in \C$ may be written as $\vc=\Psi\left( \va, \vb\right),$ where $\va\in \C_{1}, \vb\in \C_{2}$. According to (\ref{eq:mapping}),
the number of $\mathtt{G,C}$ symbols in $\vc$ equals the number of  ones in $\va$. Since $\va$ is balanced, exactly half of the symbols in $\vc$ are $\mathtt{G}$s and $\mathtt{C}$s. This implies that $\C$ has balanced GC content.

\item We prove the result by contradiction. Suppose that $\C$ is not a $\kappa$-WMU code while $\C_{1}$ is a $\kappa$-WMU code. Then, there exist sequences $\vc, \vc' \in \C$ such that a proper prefix of $\vc$ of length at least $\kappa$ appears as a suffix of $\vc^{\prime}$. Alternatively, there exist sequences $\vp, \vc_{0}, \vc_{0}^{\prime}$ such that $\vc= \vp \vc_{0} , \vc^{\prime}= \vc_{0}^{\prime} \vp$ and the length of $\vp$ is at least $\kappa$.
Next, we use the fact $\Psi$ is a bijection and find binary strings
$\va, \vb, \va_{0}, \vb_{0}$ such that 
$$\vp=\Psi\left( \va, \vb\right), \vc_{0}=\Psi\left( \va_{0}, \vb_{0}\right), \vc_{0}^{\prime}=\Psi\left( \va_{0}^{\prime}, \vb_{0}^{\prime}\right).$$
Therefore,
\begin{align*}
\vc=\vp \vc_{0}= \Psi\left( \va, \vb\right) \Psi\left( \va_{0}, \vb_{0}\right) = \Psi\left( \va \va_{0}, \vb \vb_{0} \right),\\
\vc'=\vc'_{0} \vp= \Psi\left( \va'_{0}, \vb'_{0}\right) \Psi\left( \va, \vb\right)  = \Psi\left(  \va'_{0} \va,  \vb'_{0} \vb \right),
\end{align*}
where $\va \va_{0}, \va'_{0} \va \in \C_{1}$. This implies that the string $\va$ of length at least $\kappa$ appears both as a proper prefix and suffix of two not necessarily distinct elements of $\C_{1}$. This contradicts the assumption that $\C_{1}$ is a $\kappa$-WMU code. The same argument may be used for the case that $\C_{2}$ is a $\kappa$-WMU code. 

\item For any two distinct sequences $\vc,\vc^{\prime}\in \C$ there exist $\va,\va^{\prime}\in \C_{1}, \vb, \vb^{\prime}\in \C_{2}$
such that $\vc=\Psi\left( \va, \vb\right), \vc^{\prime}=\Psi\left( \va^{\prime}, \vb^{\prime}\right)$.
The Hamming distance between $\vc, \vc^{\prime}$ equals
\begin{align*}
\sum_{1\leq i\leq n}\mathds{1}\left(c_{i}\neq c_{i}^{\prime}\right) & =\sum_{1\leq i\leq n}\mathds{1}\left(a_{i}\neq a_{i}^{\prime}\vee b_{i}\neq b_{i}^{\prime}\right)\\
 & \geq\begin{cases}
d_{1} & \textrm{if } \va\neq \va^{\prime}\\
d_{2} & \textrm{if } \vb\neq \vb^{\prime}
\end{cases} \geq \min\left(d_{1},d_{2}\right).
\end{align*}
This proves the claimed result.
\item By combining (\ref{eq:comp_2}), (\ref{eq:comp_4}) and (\ref{eq:mapping}), one can easily verify 
that $\overline{ \Psi ( \va, \vb) } = \Psi\left( \va, \overline{\vb}\right)$. We again prove the result by contradiction. 
Suppose that $\C$ is not an $f$-APD code. Then, there exist $\vc,\vc^{\prime}\in \C,\va,\va^{\prime}\in \C_{1}, \vb, \vb^{\prime}\in \C_{2}$ such that $\vc=\Psi\left( \va, \vb\right), \vc^{\prime}=\Psi\left( \va^{\prime}, \vb^{\prime}\right)$ 
and $\overline{\vc}_{i}^{f+i-1} = (\vc^{\prime})_{j}^{f+j-1} \textrm{ or } (\vc^{\prime})_{f+j-1}^{j}$, for some $1\leq i, j \leq n+1-f$. 
This implies that $\overline{\vb}_{i}^{f+i-1} = (\vb^{\prime})_{j}^{f+j-1} \textrm{ or } (\vb^{\prime})_{f+j-1}^{j}$, which contradicts the assumption that $\C_{2}$ is an $f$-APD code.
\end{enumerate}
\end{IEEEproof}

In the next sections, we devote our attention to establishing bounds on the size of WMU codes with error-correction, balancing and primer dimer constraints, and to devising constructions that use the decoupling principle or more specialized methods that produce larger codebooks. As the codes $\C_{1}$ and $\C_{2}$ in the decoupled construction have to satisfy two or more
properties in order to accommodate all required constraints, we first focus on families of binary codes that satisfy one or two primer constraints.

\section{Error-Correcting WMU Codes}\label{sec:wmuerror}

The decoupled binary code construction result outlined in the previous section indicates that in order to construct
an error-correcting $\kappa$-WMU code over $\mathbb{F}_4$, one needs to combine a binary error-correcting $\kappa$-WMU code with a classical error-correcting code. To the best of our knowledge, no results are available on error-correcting MU or error-correcting $\kappa$-WMU codes. 

We start by establishing lower bounds on the coding rates for error-correcting WMU codes using the constrained Gilbert-Varshamov bound~\cite{gilbert1952comparison,varshamov1957estimate}. 

For $\mathbf{a}\in\mathbb{F}_{q}^{n}$ and an integer $r\geq0$, let
$\mathcal{B}_{\mathbb{F}_{q}^{n}}\left(\mathbf{a},r\right)$ denote
the Hamming sphere of radius $r$ centered around $\mathbf{a}$, i.e.,
\[
\mathcal{B}_{\mathbb{F}_{q}^{n}}\left(\mathbf{a},r\right)=\left\{ \mathbf{b}\in\mathbb{F}_{q}^{n}\mid \, d_{H}\left(\mathbf{a},\mathbf{b}\right)\leq r\right\}, 
\]
where, as before, $d_H$ denotes the Hamming distance. Clearly, the cardinality of 
$\mathcal{B}_{\mathbb{F}_{q}^{n}}\left(\mathbf{a},r\right)$ equals 
$$\mathcal{V}_q(n,r)=\sum_{i=0}^{r}\left(\begin{array}{c} n\\
i \end{array}\right)\left(q-1\right)^{i},$$ 
independent on the choice of the center of the sphere. 
For the constrained version of Gilbert-Varshamov bound, let $X\subseteq\mathbb{F}_{q}^{n}$
denote an arbitrary subset of $\mathbb{F}_{q}^{n}$. For a sequence $\mathbf{a}\in X$, define the Hamming ball of radius $r$ in
$X$ by
\[
\mathcal{B}_{X}\left(\mathbf{a},r\right)=\mathcal{B}_{\mathbb{F}_{q}^{n}}\left(\mathbf{a},r\right)\cap X.
\]
The volumes of the spheres in $X$ may depend on the choice of $\mathbf{a}\in X$. Of interest is the 
maximum volume of the spheres of radius $r$ in $X$, 
\[
\mathcal{V}_{X,\max}\left(r\right)=\max_{\mathbf{a}\in X}\left|\mathcal{B}_{X}\left(\mathbf{a},r\right)\right|.
\]
The constrained version of the GV bound asserts that there exists a code of length $n$ over $X$, with minimum Hamming distance $d$ that
contains 
\[
M\geq\frac{\left|X\right|}{\mathcal{V}_{X,\max}\left(d-1\right)}
\]
codesequences. Based on the constrained GV bound, we can establish the following lower bound for error-correcting WMU codes. The key idea is to use a $\kappa$-WMU subset as the ground set $X\subseteq\mathbb{F}_{q}^{n}$.

\begin{thm}
\label{thm:avg_bound}(Lower bound on the maximum size of error-correcting WMU codes.) Let $\kappa$ and $n$ be two integers such that $n-\kappa-1\geq2 \ell$, for $\ell = \left\lceil \log_{q}2\left(n-\kappa+1\right)\right\rceil, q \in \{2, 4\}$. Then there exists a $\kappa$-WMU code $\mathcal{C}\subseteq\mathbb{F}_{q}^{n}$ with minimum Hamming distance $d$ and cardinality 
\begin{align}
\left|\mathcal{C}\right|\geq c_{q}\frac{q^{n}}{\left(n-\kappa+1\right)\left(\mathcal{L}_{0}-\mathcal{L}_{1}-\mathcal{L}_{2}\right)}, \label{eq:C_3_bound}
\end{align}
where
\begin{align*}
c_{q} = & \frac{\left(q-1\right)^{2}\left(2q-1\right)}{4q^{4}},
\end{align*}
and $\mathcal{L}_{0},\mathcal{L}_{1},$ and $\mathcal{L}_{2}$, are given by 
\begin{align}
\label{eq:l_0}
\mathcal{L}_{0}= & \mathcal{V}_{q}\left(n-\ell-1,d-1\right)+\left(q-2\right)\mathcal{V}_{q}\left(n-\ell-1,d-2\right), \notag
\end{align}
\begin{align}
\mathcal{L}_{1}= & \left(q-1\right)\sum_{i=\ell+2}^{n-\kappa-\ell+1}\left[\sum_{j=0}^{i-\ell-2}\left(\left(\begin{array}{c}
i-\ell-2\\
j
\end{array}\right)\left(q-2\right)^{j}\right.\right.\nonumber \\
 & \left.\left.\times\mathcal{V}_{q}\left(n-i-\ell+1,d-\ell-j-2\right)\right)\right], \notag
\end{align}
and
\begin{align}
\mathcal{L}_{2}= & \sum_{i=0}^{n-\kappa-\ell}\left(\begin{array}{c}
n-\kappa-\ell\\
i
\end{array}\right)\left(q-2\right)^{i}\mathcal{V}_{q}\left(\kappa-1,d-i-2\right). \notag
\end{align}
\end{thm}

\begin{IEEEproof} 
Assume that $X$ is a $\kappa$-WMU code over $\mathbb{F}_{q}^{n}$ generated according to 
Construction \ref{cons:WMU_q_n} and such that it has the cardinality at least $c_q \frac{q^{n}}{n-\kappa+1}$. Recall that in this case, 
$X$ is the set of sequences $\mathbf{a}\in\mathbb{F}_{q}^{n}$ that start with $\ell=\left\lceil \log_{q}2\left(n-\kappa+1\right)\right\rceil $
consecutive zeros ($\mathbf{a}_{1}^{\ell}=0^{\ell}$), $a_{\ell+1},a_{n-\kappa+1}\neq0$,
and no $\ell$ consecutive zeros appears as a subsequence in $\mathbf{a}_{\ell+2}^{n-\kappa}$.
With every $\mathbf{a}\in X$, we associate two sets $\mathcal{X}\left(\mathbf{a},d-1\right)$
and $\mathcal{Y}\left(\mathbf{a},d-1\right)$: The set $\mathcal{X}\left(\mathbf{a},d-1\right)$ includes sequences
$\mathbf{b}\in\mathbb{F}_{q}^{n}$ that satisfy the following three conditions:
\begin{itemize}
\item Sequence $\mathbf{b}$ starts with $\ell$ consecutive zeros, i.e., $\mathbf{b}_{1}^{\ell}=0^{\ell}$.
\item One has $b_{\ell+1}\neq 0$. 
\item Sequence $\mathbf{b}$ satisfies $d_{H}\left(\mathbf{a}_{\ell+1}^{n},\mathbf{b}_{\ell+1}^{n}\right)\leq d-1$.
\end{itemize}
The set $\mathcal{Y}\left(\mathbf{a},d-1\right) \subseteq \mathcal{X}\left(\mathbf{a},d-1\right)$ is the collection
of sequences $\mathbf{b}$ that contain $0^\ell$ as a subsequence in $\mathbf{b}_{\ell+2}^{n-\kappa}$,
or that satisfy $b_{n-\kappa+1}=0$. Therefore,  
\[
\mathcal{B}_{X}\left(\mathbf{a},d-1\right)=\mathcal{X}\left(\mathbf{a},d-1\right) / \mathcal{Y}\left(\mathbf{a},d-1\right),
\]
and
\[
|\mathcal{B}_{X}\left(\mathbf{a},d-1\right)|=|\mathcal{X}\left(\mathbf{a},d-1\right)| - |\mathcal{Y}\left(\mathbf{a},d-1\right)|.
\]
Let $\mathcal{L}_{0}=\left|\mathcal{X}\left(\mathbf{a},d-1\right)\right|$. Thus, 
\begin{align}
\mathcal{L}_{0}= & \mathcal{V}_{q}\left(n-\ell-1,d-1\right)+\left(q-2\right)\mathcal{V}_{q}\left(n-\ell-1,d-2\right).
\label{eq:l_0_2}
\end{align}
The result holds as the first term on the right hand side of the equation counts the number of 
sequences $\mathbf{b}\in\mathcal{X}\left(\mathbf{a},d-1\right)$
that satisfy $b_{\ell+1}=a_{\ell+1}$, while the second term counts those sequences for which $b_{\ell+1}\neq a_{\ell+1}$. 

We determine next $|\mathcal{Y}\left(\mathbf{a},d-1\right)|$. For this purpose, we look into two disjoints subsets $\mathcal{Y}^{\textrm{I}}$ and $\mathcal{Y}^{\textrm{II}}$
in $\mathcal{Y}\left(\mathbf{a},d-1\right)$ which allow us to use $\left|\mathcal{Y}\left(\mathbf{a},d-1\right)\right|\geq\left|\mathcal{Y}^{\textrm{I}}\right|+\left|\mathcal{Y}^{\textrm{II}}\right|$ and establish a lower bound on the cardinality sought.

The set $\mathcal{Y}^{\textrm{I}}$ is defined according to
\[
\mathcal{Y}^{\textrm{I}}=\bigcup_{i=\ell+2}^{n-\kappa-\ell+1}\mathcal{Y}^{\textrm{I}}\left(i\right),
\]
where $\mathcal{Y}^{\textrm{I}}\left(i\right)$ is the set of
sequences $\mathbf{b}\in\mathcal{X}\left(\mathbf{a},d-1\right)$ that
satisfy the following constraints:
\begin{itemize}
\item The sequence $\mathbf{b}$ contains the substring $0^\ell$ starting at position $i$,
i.e., $\mathbf{b}_{i}^{i+\ell-1}=0^{\ell}$.
\item It holds that $b_{i-1}\neq0$.
\item The sequence $0^\ell$ does not appears as a substring in $\mathbf{b}_{\ell+2}^{i-2}$.
\item One has $d_{\textrm{H}}\left(\mathbf{a}_{\ell+1}^{i-1},\mathbf{b}_{\ell+1}^{i-1}\right)+d_{\textrm{H}}\left(\mathbf{a}_{i+\ell}^{n},\mathbf{b}_{i+\ell}^{n}\right)\leq d-\ell-1$.
\end{itemize}
The cardinality of $\mathcal{Y}^{\textrm{I}}\left(\ell+2\right)$ can be found according to
\begin{align*}
\left|\mathcal{Y}^{\textrm{I}}\left(\ell+2\right)\right|= & \mathcal{V}_{q}\left(n-2\ell-1,d-\ell-1\right)\\
 & +\left(q-2\right)\mathcal{V}_{q}\left(n-2\ell-1,d-\ell-2\right)\\
\geq & \left(q-1\right)\mathcal{V}_{q}\left(n-2\ell-1,d-\ell-2\right)
\end{align*}
The first term on the right hand side of the above equality counts the sequences 
$\mathbf{b}\in\mathcal{Y}^{\textrm{I}}\left(\ell+2\right)$ for which $b_{\ell+1}=a_{\ell+1}$, 
while the second term counts sequences for which $b_{\ell+1}\neq a_{\ell+1}$. The inequality follows from the
fact that $\mathcal{V}_{q}\left(n-2\ell-1,d-\ell-1\right)\geq\mathcal{V}_{q}\left(n-2\ell-1,d-\ell-2\right)$.

To evaluate the remaining terms $\mathcal{Y}^{\textrm{I}}\left(i\right)$ for $\ell+3\leq i\leq n-\kappa-\ell+1$, assume that $d_{\textrm{H}}\left(\mathbf{a}_{\ell+1}^{i-2},\mathbf{b}_{\ell+1}^{i-2}\right)=j$. In this case, there are at least 
$$\left(\begin{array}{c}
i-\ell-2\\
j
\end{array}\right)\left(q-2\right)^{j}$$ 
possible choices for $\mathbf{b}_{\ell+1}^{i-2}$. This result 
easily follows from counting the number of ways to select the $j$ positions in $\mathbf{a}_{\ell+1}^{i-2}$ on which
the sequences agree and the number of choices for the remaining symbols which do not include the corresponding values
in $\mathbf{a}$ and $0$. As no additional symbol $0$ is introduced in $\mathbf{b}_{\ell+1}^{i-2}$, $\mathbf{b}_{\ell+2}^{i-2}$ does not contain the substring $0^\ell$ as $\mathbf{a}_{\ell+2}^{i-2}$ avoids that string; similarly, $b_{\ell+1}\neq0$. 

On the other hand, there are $q-1$ possibilities for $b_{i-1}\in\mathbb{F}_{q}\setminus\left\{ 0\right\}$,
and to satisfy the distance property we have to have 
$$d_{\textrm{H}}\left(\mathbf{a}_{i+\ell}^{n},\mathbf{b}_{i+\ell}^{n}\right)\leq d-\ell-j-2.$$
Therefore, 
\begin{align*}
\left|\mathcal{Y}^{\textrm{I}}\left(i\right)\right|\geq & \left(q-1\right)\left[\sum_{j=0}^{i-\ell-2}\left(\begin{array}{c}
i-\ell-2\\
j
\end{array}\right)\left(q-2\right)^{j}\right.\\
 & \left.\times\mathcal{V}_{q}\left(n-i-\ell+1,d-\ell-j-2\right)\right].
\end{align*}
Hence, the cardinality of $\mathcal{Y}^{\textrm{I}}$ may be bounded
from below as
\begin{align*}
\left|\mathcal{Y}^{\textrm{I}}\right|\geq & \mathcal{L}_{\textrm{I}},
\end{align*}
where 
\begin{align}
\mathcal{L}_{\textrm{I}}= & \left(q-1\right)\sum_{i=\ell+2}^{n-\kappa-\ell+1}\left[\sum_{j=0}^{i-\ell-2}\left(\left(\begin{array}{c}
i-\ell-2\\
j
\end{array}\right)\left(q-2\right)^{j}\right.\right.\nonumber \\
 & \left.\left.\times\mathcal{V}_{q}\left(n-i-\ell+1,d-\ell-j-2\right)\right)\right]. \label{eq:l_1_2}
\end{align}

The set $\mathcal{Y}^{\textrm{II}}$ comprises the set
of sequences in $\mathbf{b}\in\mathcal{X}\left(\mathbf{a},d-1\right)$ that have the following properties: 
\begin{itemize}
\item The sequence $0^\ell$ does not appear as a substring in $\mathbf{b}_{\ell+2}^{n-\kappa}$.
\item It holds $b_{n-\kappa+1}=0$.
\item One has 
$$d_{\textrm{H}}\left(\mathbf{a}_{\ell+1}^{n-\kappa},\mathbf{b}_{\ell+1}^{n-\kappa}\right)+d_{\textrm{H}}\left(\mathbf{a}_{n-\kappa+2}^{n},\mathbf{b}_{n-\kappa+2}^{n}\right)\leq d-2.$$
\end{itemize}
It is easy to verify that 
$$\mathcal{Y}^{\textrm{II}}\subseteq\mathcal{X}\left(\mathbf{a},d-1\right)\setminus\left[\mathcal{B}_{X}\left(\mathbf{a},d-1\right)\cup\mathcal{Y}_{\textrm{I}}\right].$$
Following the same arguments used in establishing the bound on the cardinality
of $\mathcal{Y}^{\textrm{II}}$, on can show that 
\begin{align*}
\left|\mathcal{Y}^{\textrm{II}}\right|\geq & \mathcal{L}_{2},
\end{align*}
where 
\begin{align}
\mathcal{L}_{2}= & \sum_{i=0}^{n-\kappa-\ell}\left(\begin{array}{c}
n-\kappa-\ell\\
i
\end{array}\right)\left(q-2\right)^{i}\mathcal{V}_{q}\left(\kappa-1,d-i-2\right). \label{eq:l_2}
\end{align}
As a result, for each $\mathbf{a}\in X,$ we have
\begin{align*}
\left|\mathcal{B}_{X}\left(\mathbf{a},d-1\right)\right|= & \left|\mathcal{X}\left(\mathbf{a},d-1\right)\right|-\left|\mathcal{Y}\left(\mathbf{a},d-1\right)\right|\\
\leq & \mathcal{L}_{0}-\mathcal{L}_{1}-\mathcal{L}_{2}.
\end{align*}
Note that $\mathcal{L}_{0},\mathcal{L}_{1},\mathcal{L}_{2}$ are independent
from $\mathbf{a}$. Therefore,
\begin{align*}
\mathcal{V}_{X,\max}\left(d-1\right) \, \leq \, & \, \mathcal{L}_{0}-\mathcal{L}_{1}-\mathcal{L}_{2}.
\end{align*}
This inequality, along with the constrained form of the GV bound, establishes the validity of the claimed result.
\end{IEEEproof}

\begin{figure*}[h]
    \centering
    \includegraphics[width=0.55\textwidth]{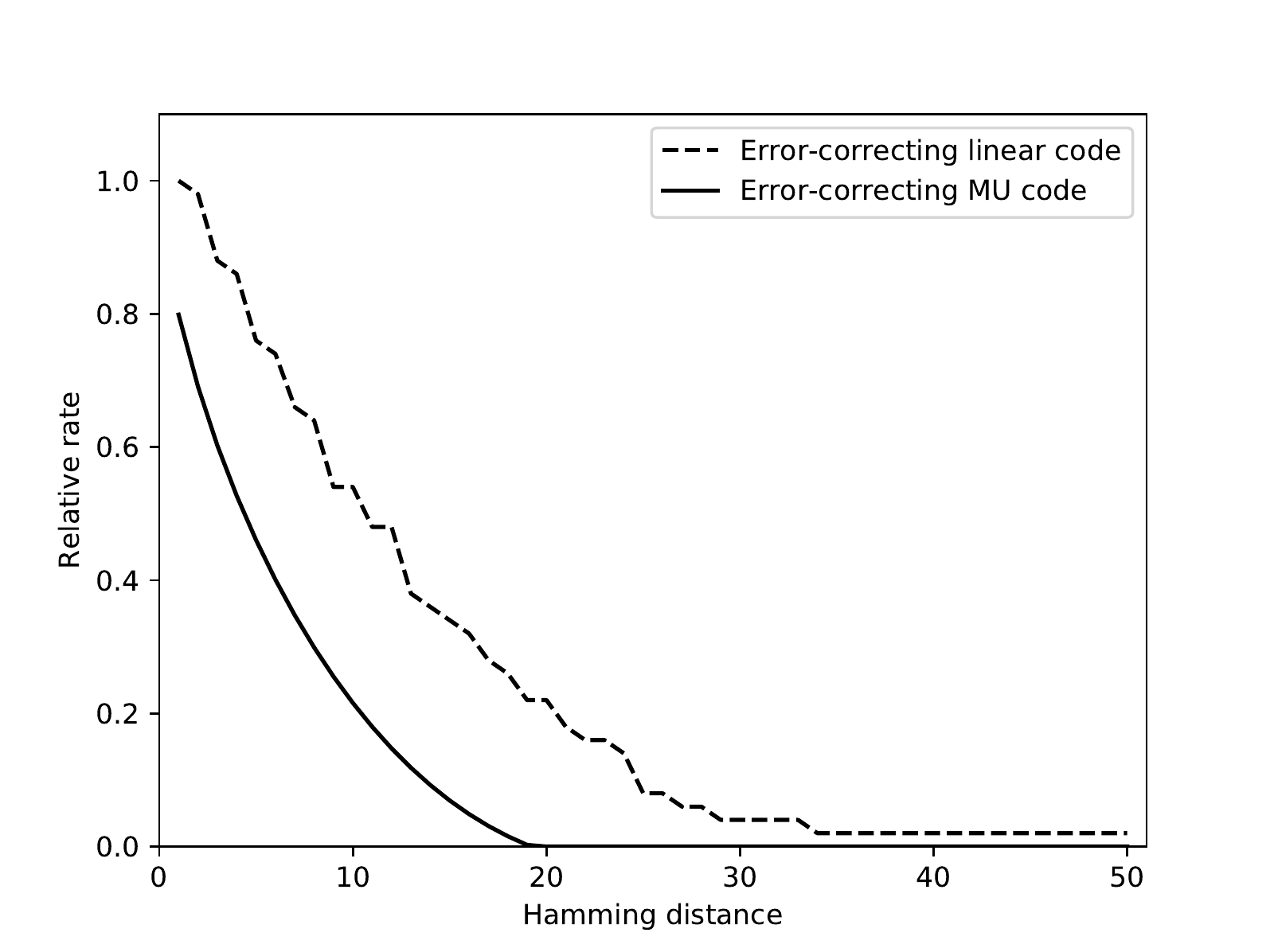}
    \caption{Comparison of two different lower bounds for binary codes: Error-correcting MU codes (inequality ~(\ref{eq:C_3_bound}) of Theorem \ref{thm:avg_bound}) and the best known linear error-correcting codes; $n=50,\kappa=1$, as $\kappa=1$-WMU codes are MU codes.}
    \label{fig:emu}
\end{figure*}

Figure \ref{fig:emu} plots the above derived lower bound  on the maximum achievable rate for error-correcting $\kappa$-WMU codes~(\ref{eq:C_3_bound}), and for comparison, the best known error-correcting linear codes for binary alphabets. The parameters used are $n=50,\kappa=1, q=2$, corresponding to MU codes.

To construct $q=4$-ary error-correcting $\kappa$-WMU codes via the decoupled construction, we need to have at our disposition an error-correcting $\kappa$-WMU binary code. In what follows, we use ideas similar to Tavares' synchronization technique~\cite{tavares1968study} to construct such codes.
We start with a simple lemma and a short justification for its validity.

\begin{lem}\label{lem:cyclic}
Let $\C$ be a cyclic code of dimension $\kappa$. Then the run of zeros in any nonzero codesequence is at most $\kappa-1$.
\end{lem}
\begin{IEEEproof}
Assume that there exists a non-zero codesequence $c(x)$, represented in polynomial form, with a run of zeroes of length $\kappa$. 
Since the code is cyclic, one may write $c(x)=a(x)g(x)$, where $a(x)$ is the information sequence corresponding to $c(x)$ and $g(x)$ is the generator polynomial of the code. Without loss of generality, one may assume that the run of zeros appears at positions $0,\ldots,\kappa-1$, so that $\sum_{i+j=s}\,a_i\,g_j=0$, for $s\in \{{0,\ldots,\kappa-1\}}$. The solution of the previous system of equations gives $a_0=a_1=\ldots=a_{\kappa-1}=0$, contradicting the assumption that $c(x)$ is non-zero.
\end{IEEEproof}
\begin{cons}($\mathtt{d_{}-HD\_\kappa-WMU\_q\_n}$ Codes)\label{cons:d-HD_k-WMU_q_n}
Construct a code $\C \subseteq \FF_{q} ^{n}$ according to
$$\C = \left\{ \va + \ve \mid \va \in \C_1, \ve=(1,0,\ldots,0) \right\}$$
where $\C_1$ is a $[n,\kappa-1,d]$ cyclic code.
\end{cons}

We argue that $\C$ is a $\kappa$-WMU code, with minimum Hamming distance $d$.
To justify the result, we first demonstrate the property of weakly mutually uncorrelatedness.
Suppose that on the contrary the code is $\C$ is not $\kappa$-WMU. 
Then there exists a proper prefix $\vp$ of length at least $\kappa$ such that 
both $\vp\va$ and $\vb \vp$ belong to $\C$.
In other words, the sequences $(\vp \va)-\ve$ and $(\vb \vp)-\ve$ belong to $\C_1$. 
Consequently, $(\vp \vb)-\ve'$ belongs to $\C_1$, where $\ve'$ is a cyclic shift of $\ve$.
Hence, by linearity of $\C_1$, $\vz\triangleq \vzero (\va-\vb)+\ve'-\ve$ belongs to $\C_1$.
Now, observe that the first coordinate of $\vz$ is one, and hence nonzero. 
But $\vz$ has a run of zeros of length at least $\kappa-1$, which is a contradiction.
Therefore, $\C$ is indeed a $\kappa$-WMU code.
Since  $\C$ is a coset of $\C_1$, the minimum Hamming distance property follows immediately.

As an example, consider the family of primitive binary $t$-error-correcting BCH codes with parameters
$[n=2^m-1,\geq n-mt, \geq 2t+1]$. The family is cyclic, and when used in Construction~\ref{cons:d-HD_k-WMU_q_n}, it results in an error-correcting $(n-mt+1)$-WMU code of minimum distance $2t+1$. 
The rate of such a code is $\frac{n-mt}{n} \geq 1-\frac{mt}{2^m - 1}$,  
while according to the Theorem \ref{thm:M_2}, the cardinality of the optimal-order corresponding $\kappa$-WMU code is at least $\frac{0.04688 \times 2^n}{mt}$, corresponding to an information rate of at least
\begin{align}
\frac{\log(\frac{0.04688 \times 2^n}{mt})}{n} > 1-\frac{5 + \log{(mt)}}{2^m - 1}.& 
\end{align}
As an illustration, we compare the rates of the BCH-based $\kappa$-WMU and the optimal $\kappa$-WMU codes for different values of 
$m = 10, t=1,3,5$:
\begin{enumerate}
\item $m=10, t = 1$: In this case our BCH code has length $1023$, dimension $1013$, and minimum Hamming distance $3$. This choice of a code results in a binary $1014$-WMU code with minimum Hamming distance $3$, and information rate $0.9902$,
while the optimal binary $1014$-WMU code has information rate greater than $0.9919$.
\item $m=10, t = 3$: In this case our BCH code has length $1023$, dimension $993$, and minimum Hamming distance $7$. This choice of a code results in a binary $994$-WMU code with minimum Hamming distance $7$, and information rate $0.9707$, while the optimal binary $994$-WMU code has information rate greater than $0.9903$.
\item $m=10, t = 5$: In this case our BCH code has length $1023$, dimension $973$, and minimum Hamming distance $11$. This choice of a code results in a binary $974$-WMU code with minimum Hamming distance $11$, and information rate $0.9511$,
while the optimal binary $974$-WMU code has information rate greater than $0.9896$.
\end{enumerate} 

Next, we present a construction for MU ECC codes of length $n$, minimum Hamming distance $2t+1$ and of size roughly $(t+1)\log n$. This construction outperforms the previous approach for codes of large rate, whenever $t$ is a small constant.

Assume that one is given a linear code of length $n'$ and minimum Hamming distance $d_H=2t+1$, equipped with a systematic encoder $\mathcal{E}_H(n',t) : \{0,1\}^\kappa \to \{0,1\}^{n'-\kappa}$ which inputs $\kappa$ information bits and outputs $n'-\kappa$ parity bits. We feed into the encoder $\mathcal{E}_H$ sequences $\bf{u} \in \{0,1\}^\kappa$ that do not contain runs of zeros of length $\ell-1$ or more, where $\ell = \log(4n')$. Let $p = \lceil \frac{\kappa}{n'-\kappa} \rceil > \ell$. 
The MU ECC codesequences are of length $n = n' + \ell + 2$, and obtained according to the constrained information sequence $\bf{u}$ as:
\begin{align*}
(0,0 \ldots,0, 1,\bf{u}_1^{p}, &\mathcal{E}_H(\bf{u})_1^1, \bf{u}_{p+1}^{2p}, \mathcal{E}_H(\bf{u})_2^2,  \\
\bf{u}_{2p+1}^{3p}, &\mathcal{E}_H(\bf{u})_3^3 \ldots, \bf{u}_{(n'-\kappa-1)p+1}^{n'}, \mathcal{E}_H(\bf{u})_{n'-\kappa}^{n'-\kappa}, 1).
\end{align*}
The codesequences start with the $0^\ell1$ substring, and are followed by sequences $\bf{u}$ interleaved with parity bits, which are inserted every $p > \ell$ positions. Notice that the effect of the inserted bits is that they can extend the lengths of existing runs of zeros in $\bf{u}$ by at most one. Since $\bf{u}$ has no runs of lengths $\ell-1$ or more this means that we do not see any runs of zeros of length $\geq \ell$ in the last $n - \ell-1$ bits of $\bf{x}$. This implies that the underlying code is MU, while the ECC properties are inherited from the initial linear code.

\section{Balanced $\kappa$-WMU Codes}\label{sec:wmubalanced}

In what follows, we focus on the analysis of balanced $\kappa$-WMU codes, and start with a review of known bounds on the number of balanced binary sequences.

Let $A_{\mathtt{d-HD\_2\_n}}$ denote the maximum size of a binary code of length $n$ and minimum Hamming distance $d$, and let $A_{\mathtt{w-CST\_d-HD\_2\_n}}$ denote the maximum cardinality of a binary code with constant weight $w$, length $n$ and even minimum Hamming distance $d_{H}$.
Clearly,
\begin{align}
&A_{\mathtt{\frac{n}{2}-CST\_2-HD\_2\_n}}=\binom{n}{\frac{n}{2}}.\notag 
\end{align}
Gyorfi~\etal \cite{gyorfi1992constructions} derived several bounds for the more general function $A_{\mathtt{w-CST\_d-HD\_2\_n}}$ based on $A_{\mathtt{d-HD\_2\_n}}$.
\begin{thm}\label{thm:JO} 
For even integer $d$, $0\leq d \leq n$, and every $w$, $ 0\leq w \leq n$,
\[
\frac{\binom{n}{w}}{2^{n-1}}A_{\mathtt{d-HD\_2\_n}} \leq A_{\mathtt{w-CST\_d-HD\_2\_n}}.
\]
\end{thm}
We present next our first construction of balanced $\kappa$-WMU codes.
\begin{cons}\label{cons:Balanced_WMU_4_n} ($\mathtt{BAL\_\kappa-WMU\_4\_n}$ Codes)
Form a code $\C\in\left\{ \mathtt{A},\mathtt{T},\mathtt{C},\mathtt{G}\right\} ^{n}$ using the decoupled 
construction with component codes $\C_1$ and $\C_2$ chosen according to the following rules: 
\begin{itemize}
\item Let $\C_{1}\subseteq \left\{0,1 \right \}^n$ be a balanced code of size equal to $A_{\mathtt{\frac{n}{2}-CST\_2-HD\_2\_n}}$.
\item Let $\C_{2 } \subseteq \left\{0,1 \right \}^n$ be a $\kappa$-WMU code; one may use Construction \ref{cons:WMU_q_n} to generate $\C_{2 }$. 
\end{itemize}
\end{cons}
\begin{lem}
\label{lem:H1}Let $\C\in\left\{ \mathtt{A},\mathtt{T},\mathtt{C},\mathtt{G}\right\} ^{n}$
denote the code generated by Construction~\ref{cons:Balanced_WMU_4_n}. Then,
\begin{enumerate}
\item $\C$ is a $\kappa$-WMU code.
\item $\C$ is balanced.
\end{enumerate}
\end{lem}
\begin{IEEEproof}
\label{proof_H1}
\begin{enumerate}
\item Since $\C_{2}$ is a $\kappa$-WMU code, property ii) of Lemma~\ref{lem:DNA_mapping_properties} ensures that 
$\C$ is also a $\kappa$-WMU code. 
\item Since $\C_{1}$ is balanced, property i) of Lemma~\ref{lem:DNA_mapping_properties} ensures that $\C$ is a balanced binary code.
\end{enumerate}
This completes the proof.
\end{IEEEproof}
We discuss next the cardinality of the code $\C$ generated by Construction~\ref{cons:Balanced_WMU_4_n}.
According to Theorem~\ref{thm:M_2}, one has $|\C_{2}| = c_2 \, \frac{2^{n}}{n-\kappa+1}$. 
In addition, $|\C_{1}| = \binom{n}{\frac{n}{2}}$.
Hence, the size of $\C$ is bounded from below by:
\begin{align*}
c_2 \, \frac{\binom{n}{\frac{n}{2}} 2^{n}}{n-\kappa+1}. \notag
\end{align*}

The following Theorem proves that both Construction \ref{cons:BAL_MU_2_n} and \ref{cons:Balanced_WMU_4_n} are order optimal, in the sense that they produce codes with cardinality within a constant factor away from the maximal achievable value.
\begin{thm}
\label{thm:BWMU_2}Let $A_{\mathtt{BAL\_\kappa-WMU\_q\_n}}$ denote the maximum
size of a balanced $\kappa$-WMU code over $\mathbb{F}_{q}^{n}$, for $n\geq2$
and $q\in\left\{ 2,4\right\} $. Then,
\begin{enumerate}
\item 
\[
c_2 \, \frac{\binom{n}{\frac{n}{2}} 2^{n}}{n-\kappa+1} \leq A_{\mathtt{BAL\_\kappa-WMU\_4\_n}} \leq \frac{\binom{n}{\frac{n}{2}} 2^{n}}{n-\kappa+1}.
\]
\item 
\[
A_{\mathtt{BAL\_\kappa-WMU\_2\_n}} \leq \frac{\binom{n}{\frac{n}{2}}}{n-\kappa+1}.
\]    
\item
\[
 \frac{\binom{n}{\frac{n}{2}}}{2(n-1)} \leq A_{\mathtt{BAL\_MU\_2\_n}} \leq \frac{\binom{n}{\frac{n}{2}}}{n}.
\]  
\end{enumerate}
\end{thm}
\begin{IEEEproof}
\label{proof_BWMU_2}
To prove the upper bounds, we use the same technique as that described in Theorem \ref{thm:M_2}.
Assume that $\mathcal{C}\subseteq\mathbb{F}_{q}^{n}$ is a balanced
$\kappa$-WMU code, for $q \in \{2,4\}$, and consider the set $X$ of pairs $\left(\mathbf{a},i\right)$
where $\mathbf{a}\in\mathbb{F}_{q}^{n},i\in\left\{1,\ldots,n\right\} $,
and the cyclic shift of the sequence $\mathbf{a}$ starting at position $i$ belongs
to $\mathcal{C}$. One may easily verify that $\left|X\right|=n\left|\mathcal{C}\right|$.
On the other hand, if $\left(\mathbf{a},i\right)\in X$, then $\mathbf{a}$
is balanced itself and there are 
$\binom{n}{\frac{n}{2}} (\frac{q}{2})^{n}$
balanced sequences to select from.
Moreover, $\left(\mathbf{a},j\right)\notin X$,
for $j\notin\left\{ i\pm1,\ldots,i\pm\left(n-\kappa\right)\right\} _{\textrm{mod }n}$
due to the $\kappa$-WMU property. Hence, for a fixed
balanced sequence $\mathbf{a}\in\mathbb{F}_{q}^{n}$, there are at most
$\left\lfloor \frac{n}{n-\kappa+1}\right\rfloor $ pairs $\left(\mathbf{a},i_{1}\right),\ldots,\left(\mathbf{a},i_{\left\lfloor \frac{n}{n-\kappa+1}\right\rfloor }\right)\in X$.
This implies that 
$$\left|X\right|\leq \frac{n \binom{n}{\frac{n}{2}} (\frac{q}{2})^{n}}{n-\kappa+1}.$$
Therefore, $\left|\mathcal{C}\right|\leq \frac{\binom{n}{\frac{n}{2}} (\frac{q}{2})^{n}}{n-\kappa+1}$.

The lower bound in (i) can be achieved through Construction~\ref{cons:Balanced_WMU_4_n}, while the lower bound in (iii) can be met using Construction~\ref{cons:BAL_MU_2_n}.
 \end{IEEEproof}
We complete our discussion by briefly pointing out how to use the balanced MU code Construction~\ref{cons:BAL_MU_2_n} to derive a balanced $\kappa$-WMU code $\C\in\left\{ \mathtt{A},\mathtt{T},\mathtt{C},\mathtt{G}\right\} ^{n}$ that has the prefix balancing property with parameter $D$. 
For this purpose, we generate $\C$ according to the balanced WMU Construction \ref{cons:Balanced_WMU_4_n}. We set $\C_2 = \left\{ 0,1 \right\}^n$ and construct $\C_1$ by concatenating $\C'_1 \subseteq \left \{ 0,1 \right \}^{\kappa-1}$ and $\C''_1 \subseteq \left \{ 0,1 \right \}^{n-\kappa+1}$. Here, $\C'_1$ is balanced and $\C''_1$ is a balanced WMU code with parameter $D$. It is easy to verify that $\C$ is a balanced $\kappa$-WMU DNA code with prefix-balancing parameter $D$ and of cardinality
\begin{align*}
|\C| = & |\C'_1| \, |\C''_1| \, |\C_2| =A(\kappa-1,2,\frac{\kappa-1}{2}) \, {\rm Dyck}(\frac{n-\kappa}{2},D) \, 2^n\\
 \sim & \frac{4^{n} \, \tan^2 \left(\frac{\pi}{D+1}\right) \, \cos^{n-\kappa} \left(\frac{\pi}{D+1} \right)}{\sqrt{2 \, \pi} \, (D+1) \, (\kappa-1)^{\frac{1}{2}}}.
\end{align*}
\section{APD-MU Codes}\label{sec:pd}
Our next goal is to provide constructions for $\kappa$-WMU codes that do not form primer dimer byproducts.

We first discuss a construction of binary MU codes with the APD property.

\begin{cons}\label{cons:APD_MU_2}($\mathtt{f-APD\_MU\_2\_n}$)
Let $n,f,\ell,p$ be positive integers such that $n = pf$ and $\ell +3 \leq \frac{f}{2}$.
Let 
$$\C = \left\{ \va_1\va_2\dots\va_{2p} \mid \va \in \C_1, \va_2,\dots,\va_{2p} \in \C_2\right\}$$
where $\C_1 \subseteq\mathbb{F}_{2}^{\frac{f}{2}}$ is the set of binary sequences $\va = (a_1,\ldots, a_{\frac{f}{2}})$ such that:
\begin{itemize}
\item The sequence $\va$ starts with $0^\ell 1$ and ends with $1$;
\item The substring $\va_{\ell+1} ^{\frac{f}{2}} $ does not contain $0^\ell$ as a substring,
\end{itemize}
and where $\C_2 \subseteq\mathbb{F}_{2}^{\frac{f}{2}}$ is the set of binary sequences $\va = (a_1,\ldots, a_{\frac{f}{2}})$ such that:
\begin{itemize}
\item The sequence $\va$ ends with $1$;
\item The sequence $\va$ contains $01^\ell 0$ as a substring.
\item The sequence $\va$ does not contain $0^{\ell}$ as a substring.
\end{itemize}
\end{cons}
\begin{lem}
\label{lem:APD_MU_2}Let $\C\in\left\{ 0,1\right\} ^{n}$
denote the code generated by Construction~\ref{cons:APD_MU_2}. Then,
\begin{enumerate}
\item $\C$ is an MU code.
\item $\C$ is an $f$-APD code.
\end{enumerate}
\end{lem}
\begin{IEEEproof}
\label{proof_H1}
The proof follows from two observations, namely
\begin{enumerate}
\item The code $\C$ satisfies the constraints described in Construction~\ref{cons:MU_q_n}, and is hence an MU code.
\item Any substring of length $f$ of any sequence in $\C$ contains an element from $\C_2$ as a substring.
Hence, any substring of length $f$ in $\C$ contains $01^\ell 0$ as a substring, and so the reverse and forward complement sequence contains $10^\ell 1$. Furthermore, no proper substring of length $f$ in $\C$ contains $10^\ell 1$ as a substring.
Hence, $\C$ is also an $f$-APD code.
\end{enumerate}
\end{IEEEproof}

Next, we use Lemma \ref{lem:avoid_string} to derive a lower bound on the size of the codes $\C_1$ and $\C_2$ in Construction~\ref{cons:APD_MU_2}, and a lower bound on the size of the code $\C$.
First, notice that
\[
|\C_1| \ge \frac{2^\frac{f}{2}}{2^{\ell+2}}\left(1-\frac{\frac{f}{2}-\ell -2}{2^\ell}\right) \ge 
\frac{2^\frac{f}{2}}{2^{\ell+2}}\left(1-\frac{f}{2^{\ell + 1}}\right),
\]
which follows from Lemma~\ref{lem:avoid_string}, with $n = \frac{f}{2} - \ell - 2, n_s = \ell, t =1$.
To bound the cardinality of $\C_2$ we define an auxiliary code $\C_3 \subseteq \{0,1\} ^ {\frac{f}{2}- \ell - 3}$ such that sequences in $\C_3$ avoid $0^{\ell -1},1^{\ell -1}$ as a substring. One can once more apply Lemma \ref{lem:avoid_string} with $n = \frac{f}{2}- \ell - 3, n_s = \ell - 1, t =2$, to obtain
\[
|\C_3| \ge \frac{2^\frac{f}{2}}{2^{\ell+3}}\left(1-\frac{4 (\frac{f}{2}-\ell -3)}{2^\ell}\right) \ge 
\frac{2^\frac{f}{2}}{2^{\ell+3}}(1-\frac{f}{2^{\ell - 1}}).
\]
Notice that by inserting $01^\ell 0$ into sequences in $\C_3$ at any of the $\frac{f}{2} - \ell -2$ allowed positions, and then appending $1$ to the newly obtained sequence, we obtain a subset of $\C_2$ of size $(\frac{f}{2} - \ell -2) |\C_3|$. Therefore,
\[
|\C_2| \ge \left(\frac{f}{2} - \ell -2\right) \frac{2^\frac{f}{2}}{2^{\ell+3}}(1-\frac{f}{2^{\ell - 1}}).
\]
For $\ell = \left\lceil \log_2 (3f) \right\rceil$, one can verify that the size of the code $\C_1$ is within a constant factor of $\frac{2^{\frac{f}{1}}}{f}$, and the size of $\C_2$ is within a constant factor of $2^{\frac{f}{2}}$.
In the last step, we use the fact that $|\C| =|\C_1||\C_2|^{2p-1}$ to show that $|\C|$ is within a constant factor of $\frac{2^n}{n}$. Therefore, Construction~\ref{cons:APD_MU_2} produces an order-optimal $f$-APD MU binary codes. The result is summarized in the following theorem.

\begin{thm}
Let $A_{\mathtt{f-APD\_MU\_2\_n}}$ denote the maximum
size of an $\mathtt{f-APD\_MU\_2\_n}$ code, for positive integers
$n=pf$ such that $p$ is a constant factor. Then, there exist constants $c_3 > 0$ such that
\[
c_3\frac{2^{n}}{n}\leq A_{\mathtt{f-APD\_MU\_2\_n}} {\leq} \frac{2^{n}}{n}.
\]
\end{thm}
\begin{IEEEproof}
The lower bound is a direct consequence of Construction~\ref{cons:APD_MU_2}, while the upper bound follows from Theorem~\ref{thm:M1}, and the fact that any $\mathtt{f-APD\_MU\_2\_n}$ code is also an $\mathtt{MU\_2\_n}$ code.  
\end{IEEEproof}

\section{APD, Balanced, Error-Correcting and WMU Codes}\label{sec:all}
In what follows, we describe the main results of our work: Constructions of APD, balanced, error-correcting $\kappa$-WMU codes. 
The gist of our approach is to use the decoupling principle along with a pair of binary codes that satisfy one or two of the 
desired binary primer constraints in order to obtain a large set of proper address/primer sequences. In addition to constructions
based on the decoupling procedure, we introduce a number of other constructions that directly produce the desired $q$-ary 
codes with large codebooks, or allow for simple encoding and decoding.

Recall Construction~\ref{cons:d-HD_k-WMU_q_n}, which we showed results in an error-correcting $\kappa$-WMU DNA code. Map the elements in 
$\FF_4$ to $\{{\tt A,T,C,G}\}$ according to:
\[0\mapsto {\tt A},\ 1\mapsto {\tt C},\ \omega\mapsto {\tt T},\ \omega+1\mapsto {\tt G},\ \]
where $\omega$ is a primitive element of the field $\FF_4$.

Let $\va$ be a sequence of length $n$. Then it is straightforward to see that the sequence $(\va,\va+1^n)$ is balanced, for $\textbf{1}^n = (1,\dots,1)$. These observations lead to the simple primer construction described next.

\begin{cons}\label{cons:V1_BAL_ECC_WMU_4_n}($\mathtt{V1:BAL\_2d-HD\_\kappa-WMU\_4\_n}$ Codes)
Let $\C$ be an $[\frac{n}{2},\kappa-1,d]$ cyclic code over $\FF_4$ that contains the all ones vector $\vone$.
Then 
\[\{(\vc+\ve,\vc+\textbf{1}^\frac{n}{2}+\ve): \vc\in\C\}\]
is a GC balanced, $\kappa$-WMU code with minimum Hamming distance $2d$.
\end{cons}

The next construction follows by invoking the decoupling principle with binary error-correcting WMU codes constructed in Section~\ref{sec:wmuerror} and codes meeting the bound of Theorem~\ref{thm:JO}.

\begin{cons}\label{cons:V2_BAL_ECC_WMU_4_n}($\mathtt{V2:BAL\_d-HD\_\kappa-WMU\_4\_n}$ Codes)
 Construct a code $\C\in\left\{ \mathtt{A},\mathtt{T},\mathtt{C},\mathtt{G}\right\} ^{n}$ via the decoupled construction of Lemma~\ref{lem:DNA_mapping_properties} involving two codes:
\begin{enumerate}
\item 
A balanced code $\C_{1}$ of length $n$, with minimum Hamming distance $d$ and of size $A_{\mathtt{\frac{n}{2}-CST\_d-HD\_2\_n}}$. 
\item 
A $\kappa$-WMU code $\C_{2} \subseteq \left \{ 0,1 \right \} ^{n}$ of length $n$ and minimum Hamming distance $d$, described in Section~\ref{sec:wmuerror}.
\end{enumerate}
\end{cons}
\begin{lem}
\label{lem:H3}Let $\C\in\left\{ \mathtt{A},\mathtt{T},\mathtt{C},\mathtt{G}\right\} ^{n}$
denote the code generated by Construction~\ref{cons:V2_BAL_ECC_WMU_4_n}. Then,
\begin{enumerate}
\item $\C$ is a $\kappa$-WMU code.
\item $\C$ is balanced.
\item The minimum Hamming distance of $\C$ is at least $d$.
\end{enumerate}
\end{lem}
\begin{exa}\label{cor:H3}
The size of the code $\C$ obtained from Construction~\ref{cons:V2_BAL_ECC_WMU_4_n} equals
\begin{align}
|\C|= & |\C_{1}|\, |\C_{2}| \nonumber\\
\geq&  \, c_2 \, \frac{2^n \, A_{\mathtt{\frac{n}{2}-CST\_d-HD\_2\_n}}}{ (n-\kappa+1) \, (\mathcal{L}_{0}-\mathcal{L}_{1}-\mathcal{L}_{2})} \nonumber\\
\geq&  \, 0.09376\, \frac{\binom{n}{\frac{n}{2}} \, A_{\mathtt{d-HD\_2\_n}}}{(n-\kappa+1) \, (\mathcal{L}_{0}-\mathcal{L}_{1}-\mathcal{L}_{2})}\label{eq:BEWMU}.
\end{align}
The last two inequalities follow from the lower bounds of Lemma~\ref{thm:avg_bound} and Theorem~\ref{thm:JO}, respectively.
\end{exa}
\begin{figure*}[h]
    \centering
    \includegraphics[width=0.55\textwidth]{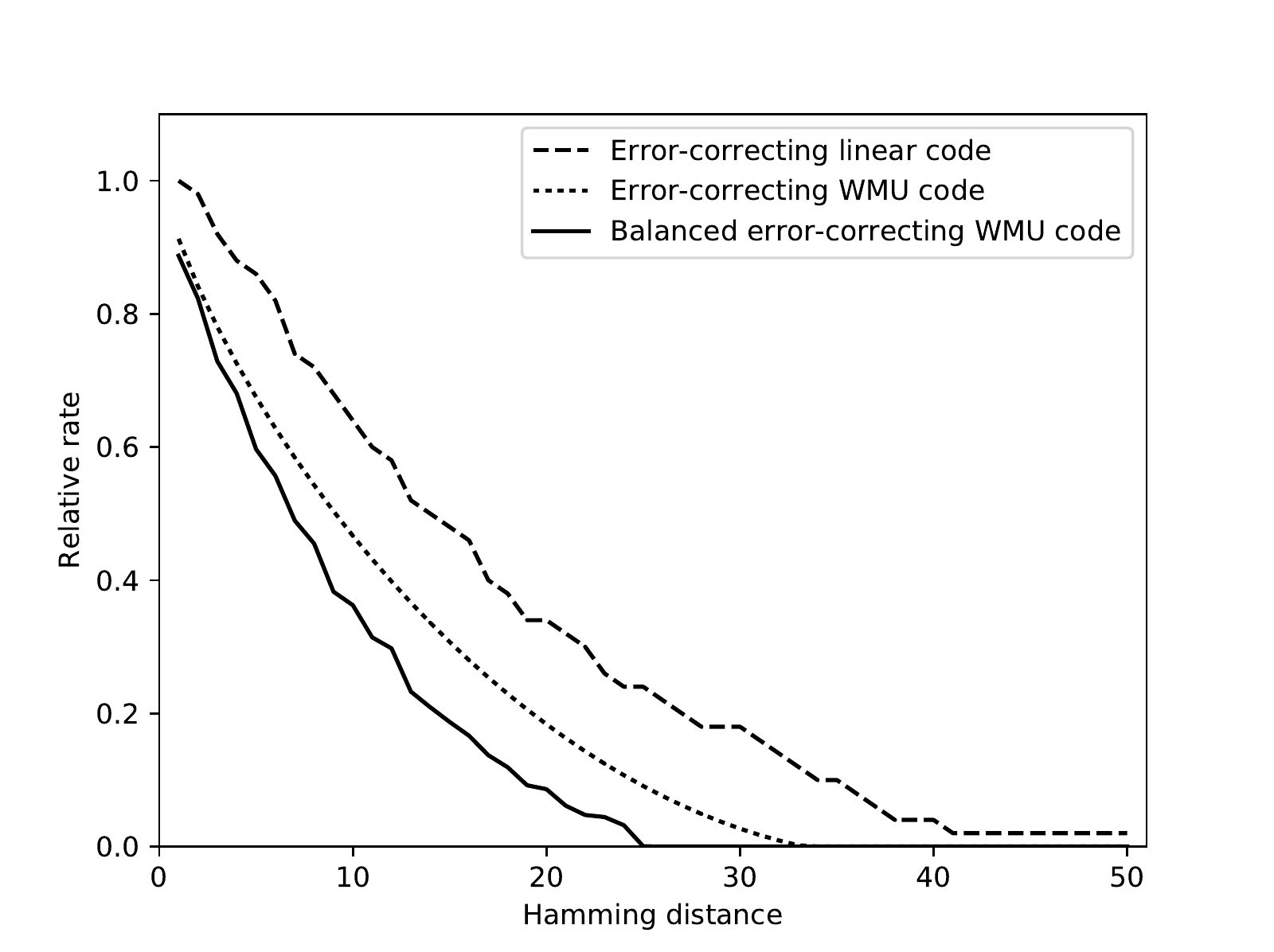}
    \caption{Comparison of three different lower bounds for quaternary codes: Balanced error-correcting $\kappa$-WMU codes (inequality (\ref{eq:BEWMU}) in Example \ref{cor:H3}), error-correcting $\kappa$-WMU codes (inequality (\ref{eq:C_3_bound}) in Theorem~\ref{thm:avg_bound}) and the best known linear error-correcting codes; $n=50,\kappa=25$.}
    \label{fig:ebwmu}
\end{figure*}

Figure \ref{fig:ebwmu} plots the lower bound  on the maximum achievable rate for error-correcting $\kappa$-WMU codes~(\ref{eq:C_3_bound}), balanced error-correcting $\kappa$-WMU codes~(\ref{eq:BEWMU}), and for comparison, the best known linear error-correcting codes over quaternary alphabets. The parameters used are $n=50,\kappa=25$.

The next result shows that Construction~\ref{cons:APD_MU_2} may be used to devise sequences that are balanced, MU and do not form any PDs.
\begin{cons}\label{cons:APD_Balanced_MU_4}($\mathtt{f-APD\_BAL\_MU\_4\_n}$ Codes)
Using the decoupled code construction in Lemma \ref{lem:DNA_mapping_properties}, a balanced, MU code $\C \in \FF_4^n$ that avoids PDs may be obtained by choosing 
$\C_1 \subseteq\FF_2^{n}$ to be an $\mathtt{BAL\_2\_n}$ code, and $\C_2 \subseteq\FF_2^{n}$ to be an $\mathtt{f-APD\_MU\_2\_n}$ 
code.
\end{cons}
It is straightforward to see that $|\C_1| = \binom{n}{\frac{n}{2}}$ and that $|\C_2| \geq c_3 \frac{2^n}{n}$. Therefore, 
the size of $\C$ is at least $c_3 \frac{\binom{n}{\frac{n}{2}}\, 2^n}{n}$.
\begin{thm}
Let $A_{\mathtt{f-APD\_BAL\_MU\_4\_n}}$ denote the maximum
size of a $\mathtt{f-APD\_BAL\_MU\_4\_n}$ code. Then
\[
c_3\, \frac{\binom{n}{\frac{n}{2}}\, 2^n}{n} \leq A_{\mathtt{f-APD\_BAL\_MU\_4\_n}} {\leq} \, \frac{\binom{n}{\frac{n}{2}}\, 2^n}{n}. 
\]
\end{thm}
\begin{IEEEproof}
The lower bound  is the direct consequence of Construction~\ref{cons:APD_Balanced_MU_4}. 
To prove the upper bound, observe that any $\mathtt{f-APD\_BAL\_MU\_4\_n}$ code is also a valid $\mathtt{BAL\_MU\_4\_n}$ code. 
The upper bound on the cardinality of an $\mathtt{BAL\_MU\_4\_n}$ code may be obtained from the upper bound of Theorem~\ref{thm:BWMU_2}, pertaining to a $\mathtt{BAL\_WMU\_4\_n}$ code, by setting $\kappa=1$.  
\end{IEEEproof}



Next, we discuss an iterative construction based on an APD, balanced, error-correcting and $\kappa$-WMU seed code. 
\begin{cons}\label{cons:concatenate}
For a given integer $s\ge 1$, let $\C_{0}$
be a set of sequences in $\mathbb{F}_{q}^{s}$. Let 
$$\C=\left\{ \va_{1}\ldots \va_{m} \mid \va_{i}\in \C_{i}\textrm{ for }1\leq i\leq m\right\},$$ 
where the subset codes 
$\C_{1},\ldots,\C_{m}\subseteq \C_{0}$ are chosen according to:
\begin{align*}
 & \C_{1}\cap \C_{m}= \emptyset \\
\textrm{and } & (\C_{1}\cap \C_{m-1}=\emptyset) \textrm{ or } (\C_{2}\cap \C_{m}=\emptyset)\\
\vdots\\
\textrm{and } & ( \C_{1}\cap \C_{2}=\emptyset)\textrm{ or } \ldots\textrm{ or } (\C_{m-1}\cap \C_{m}=\emptyset).
\end{align*}
\end{cons}
\begin{lem}\label{lem:concatenate}
Let $\C\subseteq\FF_q^n$ be a code generated according to the Construction \ref{cons:concatenate}. Then
\begin{enumerate}
\item $\C$ is $2f$-APD if $\C_0$ is $f$-APD. 
\item $\C$ is balanced if $\C_0$ is balanced.
\item $\C$ and $\C_0$ have the same minimum Hamming distance.
\item $\C$ is $\kappa$-WMU if $\C_0$ is $\kappa$-WMU.
\end{enumerate}
\end{lem}
\begin{IEEEproof}
\begin{enumerate}
\item Any proper substring of length $2f$ of any codesequence in $\C$ contains a proper substring of length $f$ of a codesequence in $\C_0$. Then, $\C$ is $2f$-APD if $\C_0$ is $f$-APD.

\item Codesequences in $\C$ form by concatenating codesequences in $\C_0$. If $\C_0$ is balanced then each codesequences in $\C$ is also balanced.

\item Again, any two distinct codesequences in $\C$ differ in at least one of the concatenated codesequences from $\C_0$. Therefore, $\C$ and $\C_0$ have identical minimum Hamming distance.  
  

\item For any pair of not necessarily distinct $\va,\vb \in \C$ and for $\kappa\leq l < n$,
we show that $\va_{1}^{l}$ and $\vb_{n-l+1}^{n}$ cannot be identical. This establishes that the constructed concatenated code is WMU. Let $l=is+j,$ where $i=\left\lfloor \frac{l}{s}\right\rfloor$ and $0\leq j<s$. We consider three different scenarios for the index $j$:
\begin{itemize}
\item $j=0$; In this case, $1\leq i<m$. Therefore, 
$(\C_{1}\cap \C_{m-i+1}=\emptyset) \textrm{ or } \ldots\textrm{ or } (\C_{i}\cap \C_{1}=\emptyset)$
implies that $\va_{1}^{l}\neq \vb_{n-l+1}^{n}$. 
\item $0<j< \kappa$; Again, one can verify that $1\leq i<m$. It is easy
to show that $\va_{l-s+1}^{l-j}$ is a suffix of length $s-j$ of a sequence in $\C_{0}$ and 
$\vb_{n-s+1}^{n-j}$ is a prefix of length $s-j$
of an element in $\C_{0}$. Since $\kappa<s-j<s,$ one has $\va_{l-s+1}^{l-j}\neq \vb_{n-s+1}^{n-j}$.
Hence, $\va_{1}^{l}\neq \vb_{n-l+1}^{n}$.
\item $\kappa \leq j<s$; In this case, $\va_{l-j+1}^{l}$ is a proper prefix of length
$j$ of a sequence in $\C_{0},$ and $\vb_{n-j+1}^{n}$ is a proper suffix
of length $j$ of an element in $\C_{0}$. Since $\kappa\leq j<s,$ one has
$\va_{l-j+1}^{l}\neq \vb_{n-j+1}^{n}$ and $\va_{1}^{l}\neq \vb_{n-l+1}^{n}$.
\end{itemize}
\end{enumerate}
\end{IEEEproof}

\section{Information Encoding with WMU addresses}\label{sec:dna_code}

In order to store user data in DNA, one needs to encode the binary information into relatively short sequences of nucleotide, each of which is equipped with a unique address sequence that satisfies the constraints outlined in the previous section. As already described,
in order to enable accurate random access via PCR, the information bearing content of the sequences has to avoid the set of address sequences. This leads to the following problem formulation.

Given a set of sequences $\A \subseteq \mathbb{F}_{q}^{n}$, 
let $\C_{\A}(N) \subseteq \mathbb{F}_{q}^{N}$ denote 
another collection of sequences that avoid members of $\A$ as substrings, i.e., $\va \neq \vb_{i}^{n+i-1}$ for every $\va \in \A$,
$\vb \in \C_{\A}(N),$ $1 \leq i \leq N-n+1$. We refer to $\A$ as the set of addresses and $C_{\A}(N)$ as the set of address-avoiding information blocks.

We discuss next three different schemes for constructing an information codebook $\C_{\A}(N)$ of sequences of length $N$ for particular sets of address sequences $\A$.

\begin{figure*}[h]
    \centering
    \includegraphics[]{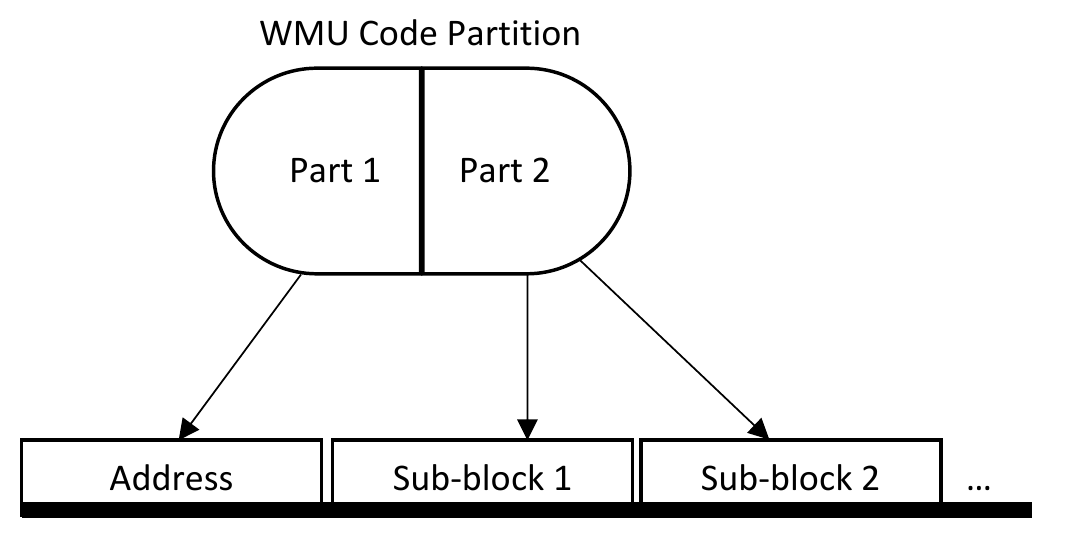}
    \caption{Concatenation construction for information blocks avoiding $\kappa$-WMU primer sequences. The gist of the approach is to use a subset of address
    sequences for actual addressing, and the remaining sequences as blocks to be concatenated.}
    \label{fig:concat}
\end{figure*}

Let $\C$ be an $\kappa$-WMU code over $\mathbb{F}_{q}^{n}$, for $\kappa \leq \frac{n}{2}$, and let $\A \subset \C$.
For a given integer $s \geq 1$, let $N=s \, n$ and define $\C_{\A}(N) \subseteq \mathbb{F}_{q}^{N}$ as the collection of all 
sequences $\vb\in\mathbb{F}_{q}^{N}$ of the form
\[
\vb = \vb_{1}\dots \vb_{s},
\]
where $\vb_{i}\in \C - \A$, for $1\leq i\leq s$. This construction is illustrated in Figure~\ref{fig:concat}. We show next that no $\vb \in \C_{\A}(N)$ contains any $\va\in \A$ as a substring. 

The proof follows by contradiction. Assume that $\va$ appears as substring in $\vb_{i}\vb_{i+1}$, for some $1\leq i < s$.
Since $\va \in \A$ and $\vb_{i},\vb_{i+1}\in \C - \A$, $\va$ may be written as $\va= \vs\vp,$ where $\vs$ is a proper suffix of $\vb_{i}$ and $\vp$ is a proper prefix of $\vb_{i+1}$. Then one of the two strings $\vp$ or $\vs$ has length greater than or equal to $\frac{n}{2} \geq \kappa$, which contradicts the fact that $\C$ is an $\kappa$-WMU code. 

The previously described construction may be easily extended to obtain error-correcting information blocks $\C_{\A}(N)$.  
We start by identifying a bijection $\mathcal{M}$ from $\C -\A$ to a finite field $\FF_{p^t}$ with appropriate prime parameter 
$p$ and $t \geq 1$; for this purpose, we expurgate codesequences from the set $\C$ so that $|\C| - |\A| = p^{t}$. 

The bijection $\mathcal{M}$ is used to convert every sequence $\vb = \vb_1 \dots \vb_s$ in $\C_{\A}(N)$ to a sequence $\vv = \vv_1 \dots \vv_s \in \FF_{p^t}^{s}$, where $\vv_i = \mathcal{M}(\vb_i),$ for $1 \leq i \leq s$. The sequence $\vv$ is encoded using an $[r,s,r-s+1]_{p^t}$ Reed-Solomon (RS) error correcting code to arrive at 
a codesequence $\vw = \vw_1 \dots \vw_r \in  \FF_{p^t}^{r}$, where $\vw_i \in \FF_{p^t}$ for $1 \leq i \leq r$. Since $\mathcal{M}$ is a bijection, one can apply $\mathcal{M}^{-1}$ to $\vw$ to reconstruct $\vc = \vc_1 \dots \vc_r \in \FF_q^{s \, r}$, where $\vc_i = \mathcal{M}^{-1}(\vw_i),$ for $1 \leq i \leq r$.
Since $\vc$ is obtained by concatenating elements from $\C - \A$, it is easy to verify that $\vc$ does not contain any element of $\A$ as a substring. 
Moreover, the RS code guarantees that given a sequence $\vc$ with at most $\lfloor \frac{r-s}{2} \rfloor$ errors, one can still fully recover $\vb$. 

For the second scheme, assume that $\C_{1},\C_{2}\subseteq\mathbb{F}_{2}^{n}$
are two disjoint collections of binary sequences of length $n$ such that for all $\mathbf{a} \in\mathcal{C}_{1},$ 
the cyclic shifts of $\va$ do not belong to $\C_{2}$, i.e.,  for all $\mathbf{a} \in\mathcal{C}_{1},$ $\va_{i}^{n}\va_{1}^{i-1}\notin \C_2$
for all $1\leq i\leq n$.

Now given $\mathcal{C}_{1}$ and $\mathcal{C}_{2}$, define the set of addresses $\A \subseteq \mathbb{F}_{4}^{2n}$ as
\[
\mathcal{A}=\left\{ \Psi (\vc,\va\va)\mid\mathbf{a}\in\mathcal{C}_{1},\mathbf{c}\in \mathbb{F}_{2}^{2n}\right\} 
\]
where $\Psi$ was introduced in (\ref{eq:mapping}).
To construct $\C_{\A}(N)$, let $s \geq 1$ be an integer such that $N=s \, n$. We define $\C_{\A}(N) \subseteq \mathbb{F}_{4}^{N}$ as the collection of all 
sequences $\vb = \vb_1 \dots \vb_s \in\mathbb{F}_{4}^{N}$
where $\vb_{i} \in\mathbb{F}_{4}^{n} $ that can be written as $\vb_{i}=  \Psi (\vf_{i} , \vg_{i})$, for some $\vg_{i} \in \C_2, \vf_{i} \in \mathbb{F}_{2}^{n}$ and $1 \leq i \leq s$. We claim that $\C_{\A}(N)$ does not contain any element of $\A$ a as substring.

If $\Psi (\vc,\va\va) \in \A$ appears as a substring in a sequence $\vb \in \C_{\A}(N),$ then there exists 
an index $1 \leq i \leq s-2$ such that $\Psi (\vc,\va\va)$ is a substring of $\vb_{i} \vb_{i+2} \vb_{i+3}$.
Since $\Psi$ is a bijection one can verify that $\va \va$ appears as a substring in $\vg_{i} \vg_{i+1} \vg_{i+2}$, for $\va \in \C_1$ and $\vg_{i}, \vg_{i+1},\vg_{i+2} \in \C_2$. In addition, $\C_1  \cap \C_2 = \varnothing$ implies that $\va \va$ can be written as $\va \va = \vs \vg_{i+1} \vp$, where $\vs$ is a proper suffix of $\vg_{i}$ and $\vp$ is a proper prefix of $\vg_{i+2}$. It is clear that $\va = \vs \vp$ and $\vg_{i+1} = \vp \vs$; hence, $\vg_{i+1} \in \C_2$ is a cyclic shift of $\va \in \C_1$, which contradicts the fact that $\C_2$ contains no cyclic shifts of elements in $\C_1$. 

The last information block design scheme we present is endowed with a simple encoding and decoding procedure. Let $\A \subseteq \FF_q^n$ be a collection of sequences of length $n$ such that $\ve \notin \A$, 
where $\ve = (0,\dots,0,1)$. For the purpose of information block encoding, we may assume that $\A$ is a $\kappa$-WMU code with desired primer properties, constructed using cyclic error-correcting codes of minimum distance at least three, as described in Constructions 8 and 9. Let $N>n$ and define $\I\triangleq \{\va | \va_{1}^{n-1} \in \FF_q^{n-1},  \va_{n}^{N} \in \FF_{q-1}^{N-n+1}\}$, 
so that $|\I|=q^{n-1}(q-1)^{N-n+1}$. There is an efficient encoding scheme that maps elements of $\I$ to a set of sequences $\C_{\A}(N)$ of length $N$ that avoid all codesequences in $\A$.

Let $\vH$ be a parity-check matrix of $\A$. Hence, a sequence $\vc$ belongs to $\A$ if and only if $\vc\vH =0$. Also, since $\ve \notin \A$, one has $\ve\vH\ne 0$.
 
 To describe the encoding procedure, we define the following function $\phi:\FF_q^{n-1}\times \{0,1,\ldots,q-2\} \to \FF_q$.
 Given $\va \in\FF_q^{n-1}$ and $0\le i\le q-2$, let $a_{n}=\phi(\va,i)$ be the $i$ smallest element in $\FF_q$ such that
 $(\va,a_n)\vH\ne 0$.
 For this function to be well-defined, it suffices to demonstrate that there are at least $q-1$ elements in $\FF_q$ such that 
 appending one of them to $\va$ yields a decoding syndrome not equal to $0$. Suppose otherwise. Then there exist distinct $u,u'$ such that $(\va,u)\vH=(\va,u')\vH=0$. The last equality may be rewritten as $(\vzero,u-u')\vH=(u-u')\ve\vH=0$, contradicting the starting assumption.
 
Encoding a sequence $\va$ results in a sequence $\vb$ obtained by concatenating the following sequences:
\[
\vb_i=
\begin{cases}
\mbox{$a_i$} & \mbox{if $1\le i\le n-1$}\\
\phi(\vb_{i-n+1}^{i-1}, a_i), & \mbox{otherwise}.
\end{cases}
\]
It is straightforward to see that sequences obtained via this encoding method avoid all elements of the codebook $\A$.
\begin{table*}[h]
\centering
\caption{Summary of the optimal code constructions for various constrained WMU codes.}
\label{tab:summary}
\begin{tabular}{lllll}
Construction No.       & Name                               & Rate                  & Features              & Comment               \\ \hline
\multicolumn{1}{|l|}{\ref{cons:BAL_MU_2_n}} & \multicolumn{1}{l|}{$\mathtt{BAL\_MU\_2\_n}$} & \multicolumn{1}{l|}{$\frac{1}{2(n-1)}\binom{n}{\frac{n}{2}}$} & \multicolumn{1}{l|}{Binary, Balanced, MU} & \multicolumn{1}{l|}{} \\ \hline
\multicolumn{1}{|l|}{\ref{cons:MU_q_n}} & \multicolumn{1}{l|}{$\mathtt{MU\_q\_n}$} & \multicolumn{1}{l|}{$c_q\frac{q^n}{n}$} & \multicolumn{1}{l|}{$q$-ary, MU} & \multicolumn{1}{l|}{$q \in \{2,4\}, c_2 = 0.04688, c_4 = 0.06152$} \\ \hline
\multicolumn{1}{|l|}{\ref{cons:WMU_q_n}} & \multicolumn{1}{l|}{$\mathtt{WMU\_q\_n}$} & \multicolumn{1}{l|}{$c_q\frac{q^n}{n-\kappa+1}$} & \multicolumn{1}{l|}{$q$-ary, $\kappa$-WMU} & \multicolumn{1}{l|}{$q \in \{2,4\}, c_2 = 0.04688, c_4 = 0.06152$} \\ \hline
\multicolumn{1}{|l|}{\ref{cons:Balanced_WMU_4_n}} & \multicolumn{1}{l|}{$\mathtt{BAL\_\kappa-WMU\_4\_n}$} & \multicolumn{1}{l|}{$c_2\frac{\binom{n}{\frac{n}{2}}2^n}{n-\kappa+1}$} & \multicolumn{1}{l|}{$4$-ary, Balanced, $\kappa$-WMU} & \multicolumn{1}{l|}{$c_2 = 0.04688$} \\ \hline
\multicolumn{1}{|l|}{\ref{cons:APD_MU_2}} & \multicolumn{1}{l|}{$\mathtt{f-APD\_MU\_2\_n}$} & \multicolumn{1}{l|}{$c_3\frac{2^n}{n}$} & \multicolumn{1}{l|}{Binary, $f$-APD, MU} & \multicolumn{1}{l|}{For some constant $c_3 > 0$} \\ \hline
\multicolumn{1}{|l|}{\ref{cons:APD_Balanced_MU_4}} & \multicolumn{1}{l|}{$\mathtt{f-APD\_BAL\_MU\_4\_n}$} & \multicolumn{1}{l|}{$c_3\, \frac{\binom{n}{\frac{n}{2}}\, 2^n}{n}$} & \multicolumn{1}{l|}{$4$-ary, $f$-APD, Balanced, MU} & \multicolumn{1}{l|}{For some constant $c_3 > 0$} \\ \hline
\end{tabular}
\end{table*}

\section{Conclusions} \label{sec:conclusion}

Motivated by emerging code design problems for DNA-based data storage, we introduced the problem of address/primer sequence design. The address
design problem reduces to constructing sequences that satisfy a new form of mutual uncorrelatedness and in addition, are balanced, at sufficient Hamming distance from each 
other and such that they avoid primer dimers which arise if a substring of an address sequence is also a substring of the reverse complement sequence of the 
same or another address sequence. Our main results are listed in Table~\ref{tab:summary}. Given the constructed address sequences, we also described
information encoding methods for sequences endowed with addresses, such that they avoid any address as a proper substring. This address avoidance property allows one to randomly access desired DNA sequences via simple and inexpensive PCR reactions.

\bibliographystyle{IEEEtran}

\end{document}